\documentclass{article}
\usepackage{xcolor}
\usepackage{url}
\usepackage{graphicx}
\newcommand{\rev}[1]{{#1}}
\usepackage{amsmath}
\usepackage{amssymb}
\usepackage{amsthm}
\usepackage{array}
\usepackage{tabu}
\newtheorem{theorem}{Theorem}

\newtheorem{definition}[theorem]{Definition}

\newcommand{\fatl}{\mathfrak{L}}
\AtBeginDocument{%
  \providecommand\BibTeX{{%
    \normalfont B\kern-0.5em{\scshape i\kern-0.25em b}\kern-0.8em\TeX}}}

\makeatletter
\newcommand{\thickhline}{%
    \noalign {\ifnum 0=`}\fi \hrule height 1pt
    \futurelet \reserved@a \@xhline
}
\newcolumntype{"}{@{\hskip\tabcolsep\vrule width 1pt\hskip\tabcolsep}}
\makeatother

\begin{document}
\pagestyle{plain}

\title{Coarsening Optimization for Differentiable Programming}



\author{Xipeng Shen${}^{+\ast}$, Guoqiang Zhang${}^{+}$, Irene Dea${}^\diamond$,
Samantha Andow${}^\diamond$,\\ Emilio Arroyo-Fang${}^\diamond$, Neal Gafter${}^\diamond$, 
Johann George${}^\diamond$, Melissa Grueter${}^\diamond$, \\Erik  Meijer${}^\diamond$, 
Steffi Stumpos${}^\diamond$, Alanna Tempest${}^\diamond$, 
Christy Warden${}^\diamond$, Shannon Yang${}^\diamond$\\
{\small
${}^{+}$ North Carolina State University\hspace{.1in} 
${}^{\ast}$ CoCoPIE LLC
\hspace{.1in}
${}^\diamond$ Facebook Inc.
\hspace{.1in}}\\
{\small Contact: xshen5@ncsu.edu}
}

\maketitle

\begin{abstract}
This paper presents a novel optimization for differentiable programming named {\em coarsening optimization}. It offers a systematic way to synergize symbolic differentiation and algorithmic differentiation (AD). Through it, the granularity of the computations differentiated by each step in AD can become much larger than a single operation, and hence lead to much reduced runtime computations and data allocations in AD. To circumvent the difficulties that control flow creates to symbolic differentiation in coarsening, this work introduces {\em $\phi$-calculus}, a novel method to allow symbolic reasoning and differentiation of computations that involve branches and loops. It further avoids "expression swell" in symbolic differentiation and balance reuse and coarsening through the design of {\em reuse-centric segment of interest identification}. Experiments on a collection of real-world applications show that {\em coarsening optimization} is effective in speeding up AD, producing several times to two orders of magnitude speedups. 
\end{abstract}

\section{Introduction}
\label{sec:intro}

A program written with differentiable programming can be differentiated automatically. The differentiation results can then be used for gradient-based optimization (e.g., gradient descent) of the parameters in the program.

Differentiable programming have been used in scientific computing, physics simulations, and other domains to help mitigate the burden of manual error-prone coding of derivative computations.
Recent several years have witnessed a growing interest of differentiable programming in machine learning (ML)~\cite{Baydin+:JML2018,vanmerrienboer2019automatic} and Probabilistic Programming~\cite{BeanMachine:PMLR2020}, to accommodate the needs of various customized ML operators, user-defined operations in the learning targets (e.g., the physical environment of reinforcement learning) and statistical sampling. 

The key technique in differentiable programming is {\em automatic differentiation}. For a program ($P$) that produces output ($y$) from some given values ($X$), automatic differentiation automatically computes the derivatives (${\partial y}/{\partial x}$) ($x\in X$) without the need for users to write the differentiation code. The given program $P$ is called the {\em primal} code, and $x$ is called an {\em active input variable}.

Existing approaches of automatic differentiation fall into two categories: (i) {\em Symbolic differentiation}, which uses expression manipulation in computer algebra systems,
(ii) {\em Algorithmic differentiation}, which performs a non-standard interpretation of a given computer program by replacing the domain of the variables to incorporate derivative values and redefining the semantics of the operators to propagate derivatives per the chain rule of differential calculus (elaborated in Section~\ref{sec:background}). 

Symbolic differentiation has been commonly regarded inappropriate for differentiable programming, for several reasons: (i) It results in complex and cryptic expressions plagued with the problem of ``expression swell''~\cite{Corliss:Swell1988}. (ii) It requires models to be defined as closed-form expressions, limiting the use of control flow and other features that are common in computer programs. 

Consequently, existing differentiable programming systems are all based on {\em algorithmic differentiation (AD)}. Algorithmic differentiation computes derivatives through accumulation of values during code execution to generate numerical derivative evaluations.
In contrast with the effort involved in arranging code as closed-form expressions under the syntactic and semantic constraints of symbolic differentiation, algorithmic differentiation can be applied to regular code, allowing branching, loops, and other language features. Some examples are Autograd~\cite{autograd}, PyTorch~\cite{pytorchDiff}, JAX~\cite{jax2018github}, and Zygote~\cite{Innes:MLSys2020}.

In this work, we advocate for a hybrid approach for differentiable programming. This new approach seamlessly integrates symbolic differentiation with algorithmic differentiation through {\em coarsening}, a compiler-based technique we introduce in this work. 

The motivation of the new approach is to eliminate the large overhead in AD incurred by its fine-grained differentiation and operation overloading. Rather than differentiation at each operation, this new approach tries to enlarge the granularity to a sequence of operations, hence the name ``coarsening optimization". It identifies a part of the to-be-differentiated computations that are amenable for symbolic differentiation, elevates it to a high-level symbolic representation, applies symbolic differentiation on it, generates the code, and then integrates it back into the computation flow of AD. 

By doing that, the coarsening optimization gives four-fold benefits: (i) It avoids many calls to the fine-grained differentiation functions and the creations of many intermediate results; (ii) the symbolic representation makes it easy to directly benefit from expression simplifications by existing symbolic engines and hence leads to more efficient code being generated; (iii) it can form a synergy with computation reuse and hence amplify the benefits; (iv) it can sometimes remove the unnecessary primal computations. If what users want is only the derivative of a function, current AD still needs to run the primal computation because of the nature of its differentiation process (Sec~\ref{sec:background}). But if coarsening can be applied to the entire function, then only its generated differentiation function needs to run, foregoing the executions of the primal function.

In addition to those benefits, coarsening features several appealing properties:  (i) As the coarsening optimization typically happens at compile time, it trades a slight increase of compile time for significant runtime savings; (ii) functioning as a way to add "shortcuts" to AD, it can be seamlessly integrated into both forward and backward differentiation; (iii) it applies regardless whether the gradients are for first or higher order optimizations.

To materialize the optimization, there are several major challenges. 

{\em Challenge I:} Complexities from control flow (e.g., branches, loops). Symbolic differentiation requires a closed form of the computation, which has been regarded as difficult for code involving complex control flow. Limiting coarsening to the code segments between the appearances of such complexities in a program would result in many short code segments, leaving many optimization opportunities submerged and much power of coarsening untapped. 

{\em Challenge II:} "Expression swell". "Expression swell" is a criticism to symbolic differentiation mentioned in some literature, which refers to the observation that the derivative often has a much larger representation than the original function has~\cite{Corliss:Swell1988}. For instance, in a straightforward implementation, the derivative of the multiplication of $n$ terms becomes an expression with $n^2$ terms: $d(f_1f_2\cdots f_n)/dx = \sum_i d(f_i)/dx\prod_{j\neq i} f_j$.  

{\em Challenge III:} Tension with computation reuse. Some calculations in the primal computation may be also required in the differentiation (e.g., $e^{X\cdot\beta}$ is part of both $(1+e^{X\cdot\beta})$ and its differentiation over $\beta$, $Xe^{X\cdot\beta}$). Reuse opportunities can also exist between different parts of differentiation. The fine-grained operations in algorithmic differentiation already build on such reuses. But in coarsened differentiation, without a careful design, such reuse opportunities can get lost as the symbolic transformation reorders and reorganizes the involved calculations. On the other hand, naively maximizing computation reuse would limit the granularity of coarsening. So there is a challenge in reconciling the tension between reuse and coarsening. 




We address the challenges through two major innovations. (i) For challenge I, we introduce {\em $\phi$-calculus}, a novel method that allows symbolic reasoning and differentiation of computations with complex control flow. Building on the {\em $\phi$-function} in single static assignment (SSA), {\em $\phi$-calculus} makes the derivation of a closed form possible for computations involving complicated control flow. It further offers a set of formulae for symbolically reasoning about and differentiating the closed forms that involve $\phi$-functions. 
(ii) For challenges II and III, we propose {\em reuse-aware SOI identification} as a way to identify the code segments of interest (SOI) for coarsening. It can strike a good tradeoff between coarsening and reuse, and at the same time keep the effects of "expression swell" under control. 

\rev{Based on an AD tool for Kotlin (DiffKt), we evaluated coarsening on 18 settings of six applications on two machines.} The results show that coarsening is effective in significantly expanding the applicable scope of symbolic differentiation, and hence dramatically reducing the runtime overhead of AD. The performance improvement is substantial, 
1.03$\times$-27$\times$ speedups of the differentiation and 1.08$\times$-11$\times$ speedups of the end-to-end application execution. \rev{We further examined the potential of coarsening on several other AD tools (Zygote~\cite{Innes:MLSys2020} for Julia, Jax~\cite{jax2018github} for Python, Adept~\cite{Hogan+:TMS2014} for C++) by experimenting with the implementations of the symbolic differentiation results in their corresponding languages. The speedups from the coarsening results are even greater, 66$\times$-335$\times$, indicating the potential of coarsening in serving as a general optimization technique for AD.}

To the best of our knowledge, this is the first work that proposes a systematic approach to integrating symbolic differentiation with algorithmic differentiation for differentiable programming. The developed $\phi-$calculus offers the first method to enable symbolic differentiation of computations spanning over complex control flow. The resulting hybrid differentiation approach gets the best of both worlds, that is, the efficiency from the compile-time symbolic differentiation and the generality of AD. 

In summary, this work makes the following contributions:
\begin{itemize}
    \item It introduces {\em coarsening optimization}, the first approach to systematical integration of symbolic differentiation into algorithmic differentiation for general programs.
    \item It develops {\em $\phi$-calculus} that eliminates the barriers of control flow to symbolic differentiation.
    \item It proposes {\em reuse-aware SOI identification} to balance reuse and coarsening.
    \item It validates the benefits of coarsening, confirming its potential for significantly improving AD efficiency.
\end{itemize}

\section{Background and Terminology}
\label{sec:background}

At the foundation of AD is the chain rule. We explain it in a simple setting. Suppose $y$ is the output of a sequence computations on input $x$, and $y_i$ ($i=1,2,\cdots,k$) are the intermediate results produced during the sequence of computations from $x$ to $y$, that is, $y_1=f_1(x), y_2=f_2(y_1), \cdots, y_k=f_k(y_{k-1}), y=f(y_k)$. The chain rule says that the derivative of $y$ on $x$ (or called {\em $x$'s gradient} regarding $y$) can be computed as follows:

\[
dy/dx = dy/dy_k * dy_k/dy_{k-1} * \cdots * dy_1/dx
\]

In a program, besides the variables relevant to the deriatives of interest, there can be many other variables. To distinguish them, we call the relevant output variables like $y$ {\em active output variables}, relevant input variables like $x$ {\em active input variables}, and other relevant variables simply {\em active variables}. Further, we call the computations in the original program from active input variables to active output variables {\em primal computations}, and the computations to compute the derivatives {\em gradient computations}. 

There are two ways to interpret the chain rule, which lead to the {\em forward} and {\em backward} AD respectively. We explain them by assuming an implementation of AD via operator overloading, the most common way of implementation of AD. 

The first is to regard the rhs of the chain rule a sequence of computations from the rightmost term to the leftmost term. Corresponding to AD implementation, the derivatives of $dy_1/dx$ is computed as a side step of operator overloading when $y_1$ is computed from $x$ in the primal computation, and the result is then passed to the next step of primal computation, which computes $y_2$ and $dy_2/dy_1$ and then multiplies it with the received value of $dy_1/dx$. The process continues and produces $dy/dx$ eventually. This implementation is called {\em forward AD}. 

The second way is to regard the rhs of the chain rule a sequence of computations from the leftmost term to the rightmost term. In this case, at each step in the {\em primal} computation (which is still right-to-left), some operations needed for differentiation are recorded in a data structure, such as a stack~\cite{Hogan+:TMS2014}, as part of the operations of the overloaded operators. When the primal computations reach the last step and the gradient computation actually starts, the operations on the stack are executed in a backward order, starting from those of the leftmost term in the rhs of the chain rule. After $dy/dy_k$ is computed, the result is passed to the next step, which computes $dy_k/dy_{k-1}$ and then multiples it with the received value of $dy/dy_k$. The process continues until the gradient of $x$ is computed. This implementation is called {\em backward AD}. 

{\em Backward AD} is a more popular choice in existing AD tools because it is overall more efficient in general settings~\cite{Margossian:ADSurvey2019}. The two methods are sometimes used together.  Note that in both of them, primal computations are necessary to run so that the overloaded operators can take place, even if what the user wants are just the gradients. Coarsening optimization can lift such a requirement as shown later in this paper. 

In cases that are not differentiable (e.g., $x$=0 in $relu(x)$), AD tools approximate the gradients (e.g., using 0 at $relu(0)$); coarsening optimization preserves the same behavior.




\section{Overview of Coarsening Optimization}
\label{sec:overview}

\begin{figure}
    \centering
    \includegraphics[width=.85\textwidth]{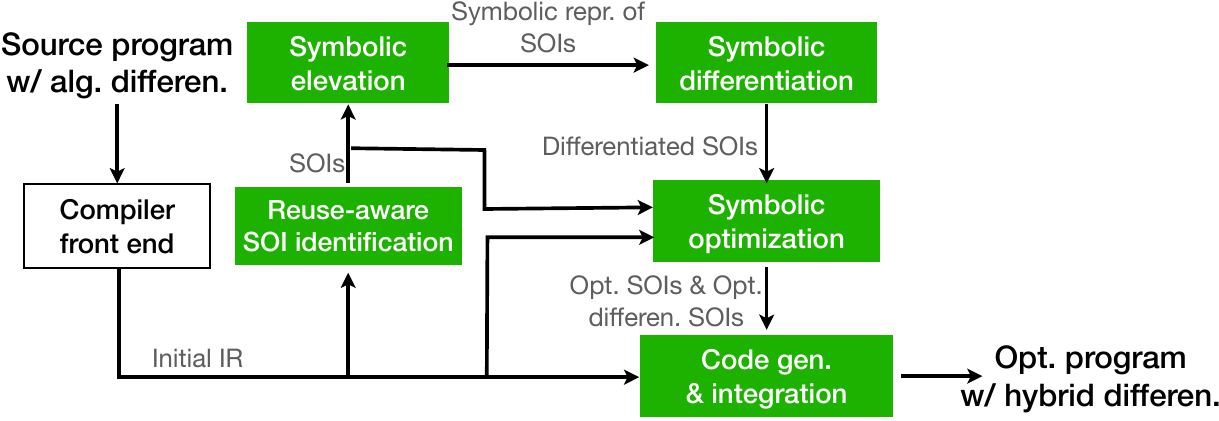}
    \caption{The overall workflow of coarsening for AD. Solid boxes are the main components in coarsening.}
    \label{fig:overview}
\end{figure}

The basic definition of {\em coarsening optimization} for AD is as follows:
\begin{definition}
Let $S$ be a sequence of program statements that implement the computations from active input $I$ to active output $P$. {\em Coarsening optimized AD} is applied to $S$ if a closed form $F$ is produced that captures the computations in $S$ and $F$ goes through a symbolic differentiation with the results integrated into the AD process.
\end{definition}

Figure~\ref{fig:overview} shows the high-level workflow of coarsening for AD. The input is a program written in a certain AD-based differentiable programming language. Coarsening works on the intermediate representation (IR) output from the front end of the default compiler. From it, the {\em reuse-aware SOI identification} component identifies the code segments of interest (SOIs), which are sent to the {\em symbolic elevation} component to produce a symbolic representation of the SOIs. The {\em symbolic differentiation} component takes them in and outputs the symbolic form of the differentiated SOIs. The {\em symbolic optimization} component optimizes both the SOIs and the differentiated SOIs while drawing on their contexts captured in the original IR. The optimizations include simplifications via algebra systems, as well as identifying the places for profitable computation reuses between SOIs and the differentiated SOIs. 

Coarsening optimization can be applied for AD at both compilation and runtime. We take compile-time optimization as the context of discussion.


\paragraph{Example} 

\begin{figure}
    \centering
    \includegraphics[width=.98\textwidth]{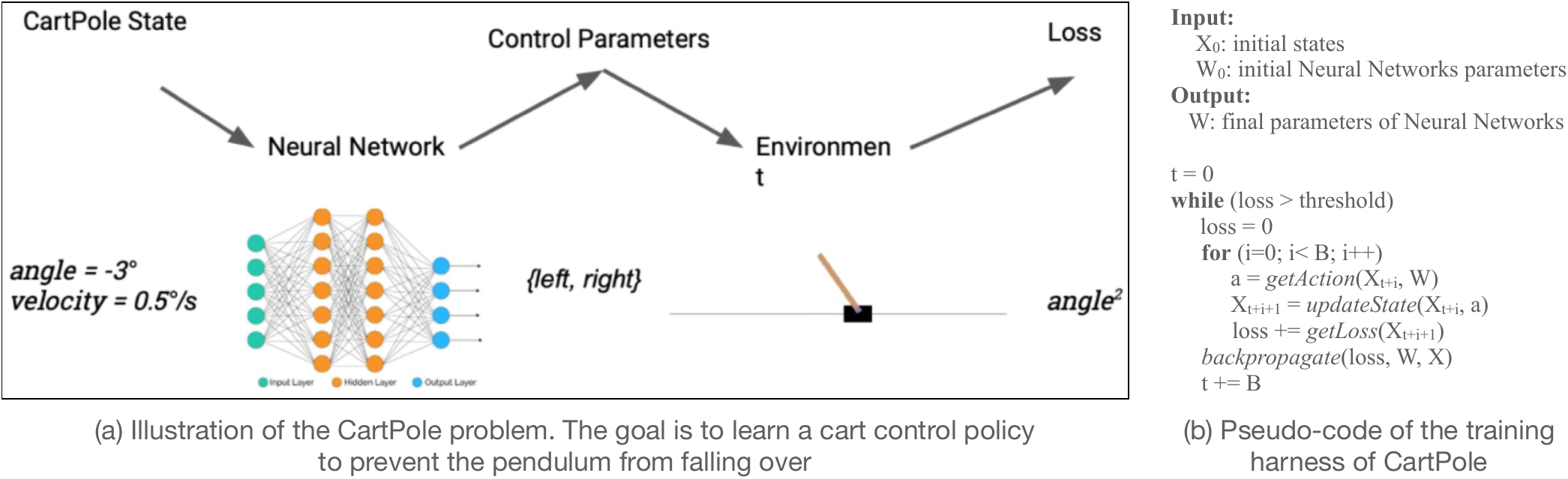}\\[4mm]
    \includegraphics[width=.98\textwidth]{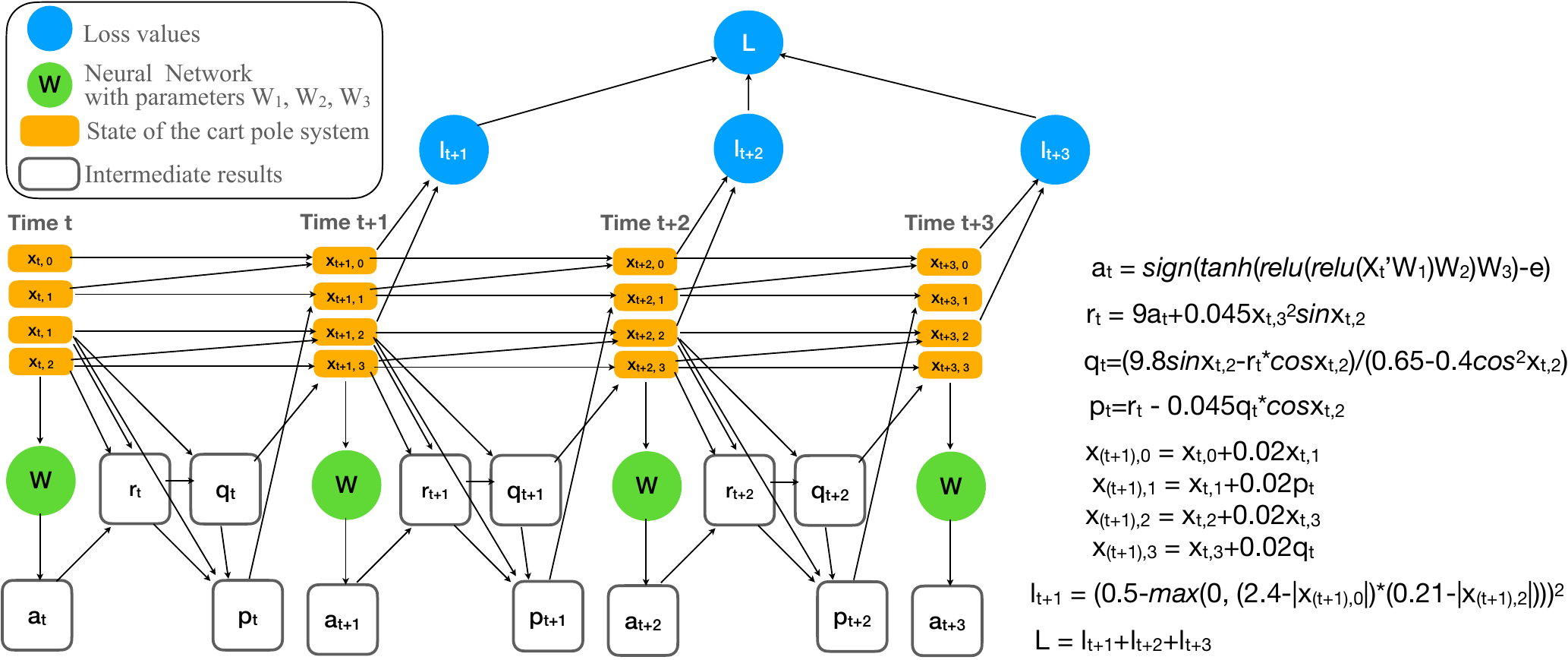}\\
    {\small (c) Computation flow of the forward pass of CartPole with the main computations shown on the right.}
    \caption{\rev{A running example CartPole. (a) Problem illustration (artwork source: fluxml.ai); (b) Pseudo-code of the training harness; (c) The simplified computation graph of the forward pass and the core computations.}}
    \label{fig:cartpole}
\end{figure}

To help convey the intuition of coarsening optimization, we use CartPole as an example. CartPole is an example that uses deep reinforcement learning (DRL). As illustrated in Figure~\ref{fig:cartpole} (a), a pole is attached by an un-actuated joint to a cart, which moves along a frictionless track. The cart is controlled by applying a force of +1 (to the right) or -1 (to the left) on the cart. The pendulum starts upright, and the goal is to learn a cart control policy to prevent the pendulum from falling over. The learning system consists of a Neural Network and a simulator of the cart and pole. At each time step, the system goes through the computation outlined in the inner loop body in Figure~\ref{fig:cartpole} (b), that is, calculating the output of the Neural Networks from the current cart and pole's states to decide the action for the cart to take, based on which, it then updates the states of the cart and pole according to the physics model. This process continues for another two time steps. \rev{The resulting total loss is then used in updating the weights of the Neural Networks via gradient descent. For each weight $w$ in the Neural Networks, (i) the program obtains the value $d_w$ for the derivative of $w$ w.r.t. the loss, (ii) $d_w$ is then used to create the differential $\lambda z. d_w * z$, and (iii) the value of weight $w$ is updated by the result of evaluating the differential w.r.t. learning rate $\eta$:
\[
w = w - (\lambda z. d_w * z)(\eta).
\]}
 The learning then continues until the Neural Networks converge. \rev{The right part of Figure~\ref{fig:cartpole} (c) shows the core computations of states update in each iteration.}


\paragraph{Intuition of Benefits}

We can get some intuition about the potential benefits of coarsening by checking the differentiation of a state at time $t+2$, $x_{t+2,0}$, over the action at time $t$, $a_t$, in Figure~\ref{fig:cartpole}. The result is needed in the computation of the gradient of $a_t$ regarding the loss $L_{t+2}$ and the total loss $L$. \rev{(The computation from the weight $W$ to $a$ is through Neural Networks, which have a standard structure; the gradient calculation of that part is through a highly polished vendor-provided library rather than the AD tool. Therefore, the actual effects of the AD on that benchmark is from the loss to $a$.)}

If we put down the entire computation from $a_t$ to $x_{t+2,0}$, we get the following (with constants already folded):

\begin{align}
x_{t+2,0} & = & x_{t,0} + 0.04x_{t,1} + 0.003636 a_t + 0.000018 {x_{t,3}}^2 sin x_{t,2} \nonumber \\ \nonumber
& & - \frac{0.00016 sin x_{t,2} - 0.00016 a_t cos^2x_{t,2} - 0.000008 {x_{t,3}}^2sin x_{t,2} cos^2x_{t,2}}{(0.65 - 0.405 cos^2x_{t,2})}\\ \label{eq:xt2}
\end{align}

It consists of 29 operations. If a (backward) AD library is used to get its differentiation over $a_t$, during the primal computation, at each of the 29 operations, a pullback function is generated for the differentiation of that operation, along with the closure and some intermediate objects allocated to hold the intermediate results that the differentiation would need to use. 

In contrast, if symbolic differentiation is applied to the expression in Equation~\ref{eq:xt2}, the result is much simpler as shown as follows. The differentiation would then need to just make an invocation to one function that consists of only several straight-line calculations. Besides saving computations, it also saves the allocations of many intermediate objects. 

\begin{align}
    \frac{d(x_{t+2,0})}{d(a_t)} = 0.003636 + \frac{0.00016cos^2x_{t,2}}{0.65-0.405cos^2x_{t,2}} \label{eq:dxt2}
\end{align}

Besides the benefits demonstrated by the CartPole example, two other benefits are worth mentioning. First, because AD libraries are typically implemented via operator overloading, they have to wrap data objects in a special type (e.g., Tensor in PyTorch) so that customized operations can be invoked during the primal computations to implement the needed AD operations. Accesses to the data objects are therefore subject to the boxing overhead. Inside the code generated by the symbolic differentiation, as no operator overloading is needed, unboxed data objects can be directly used, reducing the boxing overhead. This benefit is especially prominent when the data involved are a collection of scalars or small vectors as in many physical simulations or Probablistic Programming applications (examples in Sec~\ref{sec:evaluation}). 

The other benefit of coarsening not captured by the CartPole example is the cancellation of terms or other simplifications that symbolic transformation can often harness. We explain it with a simple expression involving two matrices ($X_1$, $X_2$) and one vector ($v$): $(X_2v)^T(X_1X_2v)$. Its symbolic differentiation over $v$ can easily combine terms with common multipliers, yielding a form $X_2^T(X_1+X_1^T)X_2v$, significantly simpler than what the default AD would compute, $X_2^T(X_1X_2v)+((X_2v)^TX_1X_2)^T$. \rev{Simplification of symbolic expressions is a common feature in symbolic engines; for more examples, please refer to the simplification module in Sympy~\cite{sympy}.}

As mentioned in Section~\ref{sec:intro}, to make {\em coarsening optimization} effective, there are three main challenges: control flows, "expression swell", and tension between coarsening and reuse. We next explain our solutions to these challenges. 



\section{Addressing Control Flow: $\phi$-Calculus}
\label{sec:phi}

Control flow complexities are commonly perceived obstacles for symbolic differentiation. For straight-line code, it is easy to derive a closed form for the computations by symbolically substituting later references with their earlier definitions in the code. In the presence of control flow branches, the complexity increases exponentially: If we build a closed form for each possible path, in the worst case, there would be $O(2^B)$ closed forms for $B$ conditional statements. That would not only increase the amount of work and execution time of symbolic differentiation, but also complicate compiler-based code generation from the differentiation results. The problem worsens when there are loops mixed with if-else. Without a closed form, symbolic differentiation cannot apply. 

Without an effective way to handle control flow, coarsening can apply only to small pieces in a program, with each piece consisting of the code between two adjacent control flow branching or merging points. For some programs, that could lead to only small optimization scopes, leaving the power of coarsening optimization untapped. 

We address the problem by proposing $\phi$-calculus. It consists of a set of notations for symbolically representing loops and conditional statements, and introduces a series of formulae to facilitate the reasoning and differentiation on the extended symbolic form. As $\phi$-calculus is inspired by the concept of $\phi$ functions in SSA, we first give a quick review of SSA.

\subsection{Background on SSA and $\phi$ Functions}

Static Single Assignment form (SSA) is a kind of code representation widely used in modern compilers~\cite{Cytron+:POPL1989,dragonBook}. Code in SSA has two properties: (i) no two static assignments assign values to the same variable; (ii) every reference refers to the value defined by a single static assignment. It uses a special $\phi$ function to resolve name ambiguities. A $\phi$ function "chooses" the right name among its (two or more) arguments based on the actual control flow. 

\begin{figure}
    \centering
    \fbox{
    \includegraphics[width=.96\textwidth]{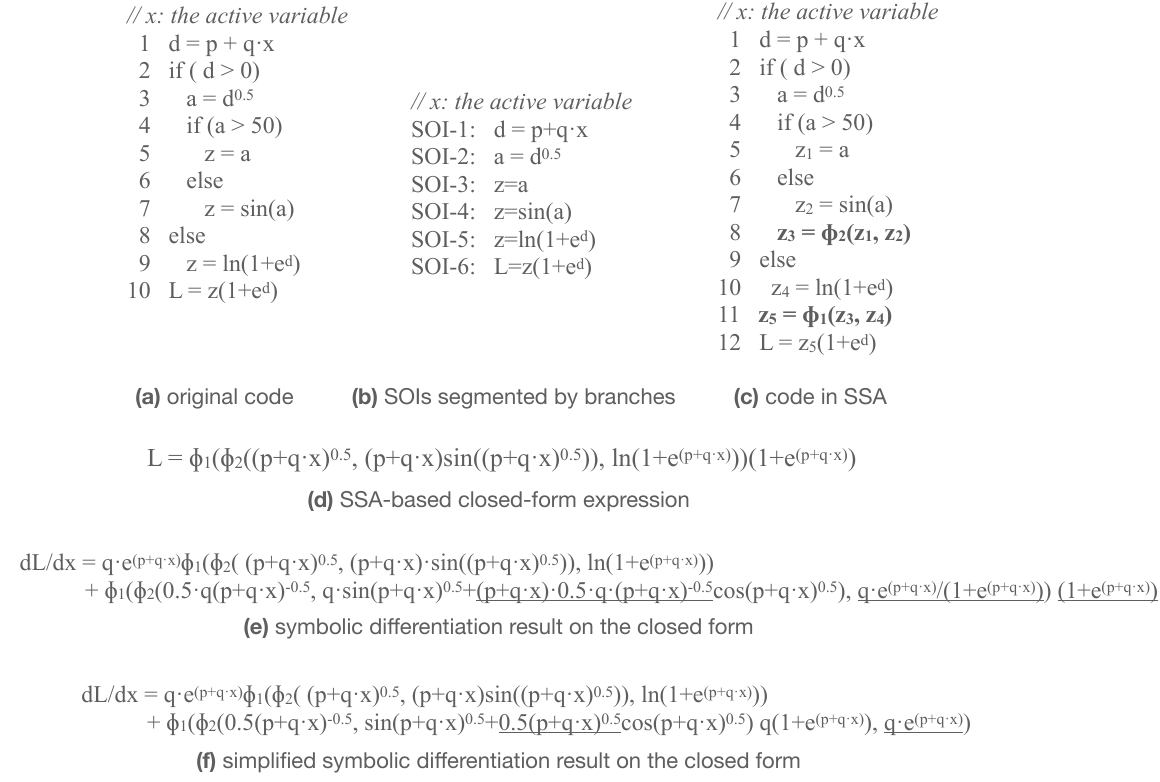}
    }
    \caption{Illustration of SSA and $\phi$-calculus on an example with conditional statements} 
    \label{fig:ifExample}
\end{figure}

Figure~\ref{fig:ifExample}(c) shows the SSA form of the code in Figure~\ref{fig:ifExample}(a). There are three assignments to $z$ in the original code. They are all replaced with different names $z_1,z_2,z_4$. Meanwhile, two $\phi$ functions are inserted in Figure~\ref{fig:ifExample}(c), with the one on line 8 resolving the name ambiguity caused by the inner if-else, and the one on line 11 for the outer if-else. 

\begin{figure}
    \centering
    \fbox{
    \includegraphics[width=.96\textwidth]{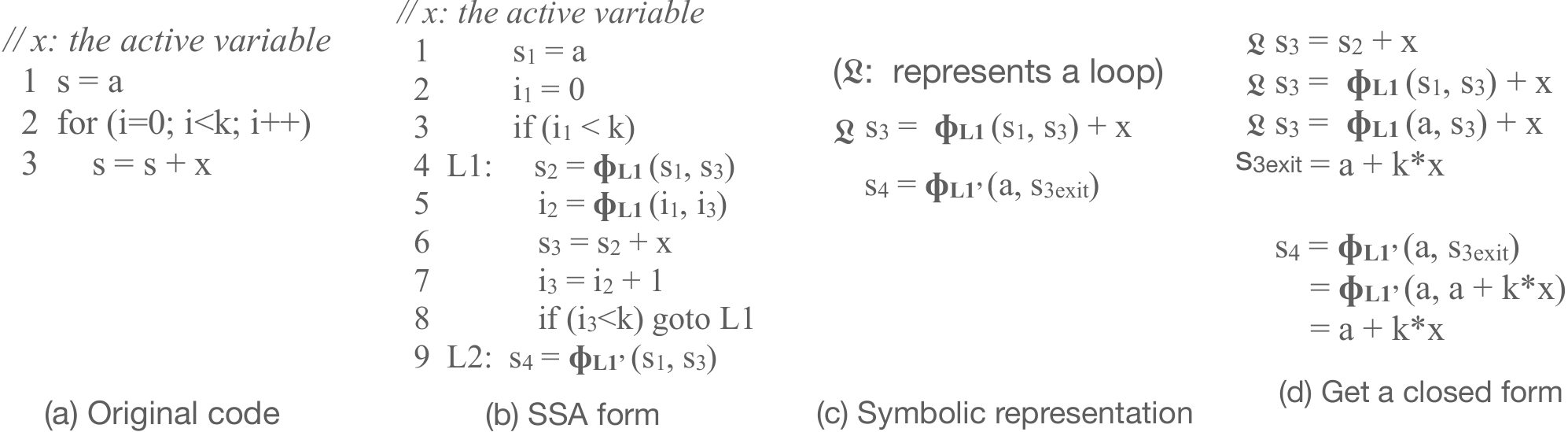}
    }
    \caption{Illustration of SSA and $\phi$-calculus on a simple loop}
    \label{fig:loopExample}
\end{figure}

Figure~\ref{fig:loopExample}(b) shows the SSA form of the loop in Figure~\ref{fig:loopExample}(a). The loop structure involves two merging points, with one at the entry (L1), the other at the exit (L2). There are two $\phi$ functions at the entry, respectively for variables $s$ and $i$; there is one $\phi$ function at the exit, for variable $s$. The former are called {\em entry $\phi$-functions} and the latter is called {\em exit $\phi$-function} of the loop~\cite{Ottenstein+:PLDI1990}. All loops, either regular or irregular (e.g., while loops with breaks), have such pairs of {\em $\phi$-functions}. 

\subsection{Notations in $\phi$-Calculus}

Inspired by SSA, $\phi$-calculus uses $\phi$ functions and other notations to symbolically represent conditional statements and loops:

\begin{itemize}
    \item $\phi(a_1, a_2, \cdots, a_k)$: the standard $\phi$ function in SSA. Numerical subscripts are sometimes added to a standard $\phi$ function to distinguish $\phi$ functions that have different conditions. 
    \item $\phi_{Li}(a_1, a_2, \cdots, a_k)$ \& $\phi_{Li'}(a_1, a_2, \cdots, a_k)$: the entry and exit $\phi$ functions of loop $i$. 
    \item ${\fatl_i}^u\;\;<S>$ : the computations in a statement $S$ are surrounded by loop $i$ with $u$ iterations. Sometimes, the loop ID ($i$) is used in $S$ to also denote the iteration number, which, by default, goes from 0 to $u-1$. \rev{In the representation of a loop for symbolic differentiation in coarsening, $u$ can be a constant, an expression, or a symbol. For irregular loops (e.g., the \texttt{while} loop in Figure~\ref{fig:bgdExample}), for instance, $u$ is a symbol in the expression that is symbolically differentiated, and its value is recorded in the execution of the primal code. In the following discussion, unless necessary, we omit $u$ and/or $i$ in the loop notations for better readability.}
    \item $\sum$, $\prod$: the standard math notations of summation and product. 
    \item ${\phi_{Li}}^{(j)}$: the instance of $\phi_{Li}$ in the $j$th iteration of loop $i$
    \item $a^{(j)}$: if $a$ is an expression in the argument list of $\phi_{Li}$, $a^{(j)}$ represents the value of $a$ after $j$ iterations of loop $i$.
    \item $a_{exit(L)}$: the value(s) of $a$ at the exit of loop $L$. 
    \item $f^{[n]}$: recursively apply function $f$ for $n$ times.
\end{itemize}

This set of notations are simple extensions of the standard $\phi$ function. But with them, code with complex control flow can now be symbolically expressed. Figure~\ref{fig:ifExample}(d), for instance, shows the symbolic form of the computation of $L$ by the code in Figure~\ref{fig:ifExample}(a). Figure~\ref{fig:loopExample}(c) shows the computation of $s$ by the loop in Figure~\ref{fig:loopExample}(a). The derivation of them involve just direct substitutions of names with their corresponding expressions; the $\phi$-notations offer ways to symbolically represent the effects of loops and conditional statements.

Note that unlike the form in Figure~\ref{fig:ifExample}(d), in Figure~\ref{fig:loopExample}(c) is not yet a closed form: There is still the presence of $\fatl$. Even for the closed form in Figure~\ref{fig:ifExample}(d), it still contains $\phi$ functions. The other component of $\phi$-calculus, {\em formulae in $\phi$-calculus}, offers the facilities for (i) getting closed forms by removing $\fatl$ and (ii) differentiating expressions involving $\phi$ functions.

\subsection{Formulae in $\phi$-Calculus}

\begin{figure}
\centering
\fbox{
\includegraphics[width=.9\textwidth]{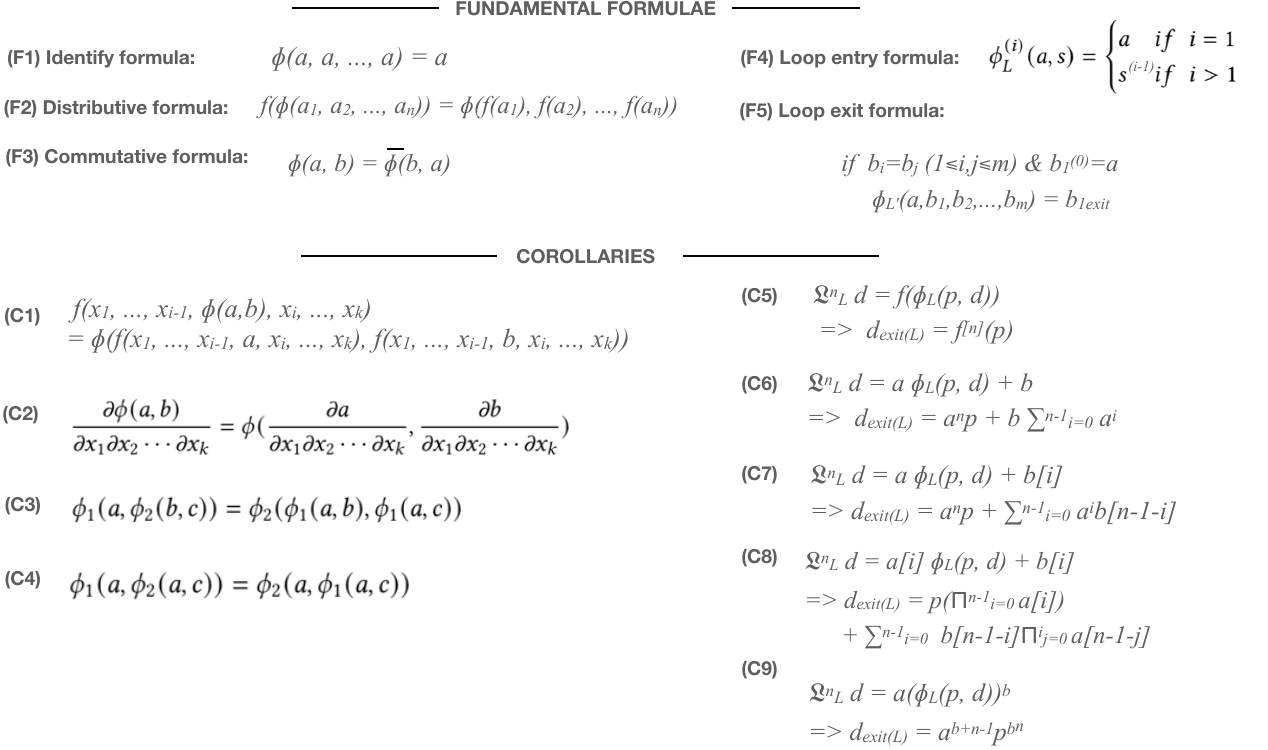}
}
\caption{Main formulae in $\phi$-calculus.}
\label{fig:formulae}
\end{figure}

Figure~\ref{fig:formulae} provides the core set of formulae in $\phi$-calculus; at the top of the figure are five fundamental formulae and at the bottom are nine useful corollaries derived from the fundamental formulae. Most of the formulae are quite straightforward, but when being used together, they are powerful in getting rid of $\phi$ and loop notations from the symbolic representation of code. We next explain each of the formulae and provide brief proofs.

\subsubsection{Fundamental $\phi$ Formulae}
\label{sec:Fundamental}

$\;$

1) {\bf Identity Formula (F1 in Figure~\ref{fig:formulae})}. The identity formula says that if all the arguments in a $\phi$ function equal to one another, the $\phi$ function can be replaced with any of its argument. It immediately follows the definition of the $\phi$ function.


2) {\bf Distributive Formula (F2 in Figure~\ref{fig:formulae})}. This formula says that a function that applies to a $\phi$ function can be distributed to each of the arguments of that $\phi$ function. It can be easily proved based on the definition of $\phi$ function. 





3) {\bf Commutative Formula (F3 in Figure~\ref{fig:formulae})}. This formula shows the relationship between a $\phi$ function and its complement. In the formula, $\overline{\phi}$ is the {\em complement} of $\phi$, that is, it chooses the first argument when $\phi$ chooses the second, and the second when $\phi$ chooses the first. The correctness of this formula immediately follows the definition.

4) {\bf Loop entry formula (F4 in Figure~\ref{fig:formulae})}. This formula shows the inherent property of a {\em loop-entry $\phi$ function}. For the definition of a {\em loop-entry $\phi$ function}, $\phi_L$ is reached always through the back edge of loop $L$ except for its first instance in that loop. The formula hence follows. This simple formula is essential for $\phi$-calculus to deal with loops as shown later.

5) {\bf Loop exit formula (F5 in Figure~\ref{fig:formulae})}. This formula says that $\phi_{L'}(a,b_1,b_2,\cdots,b_m)$ equals the value of $b_1$ at the exit of loop $L$ if (i) the value of all arguments, except the first, of $\phi_{L'}$ are the same at $\phi_{L'}$, and (ii) those arguments before the entry point of the loop have the value equaling the first argument's value $a$. Its correctness can be easily proved with the {\em identity formula}. Notice that the only time when $\phi_{L'}$ takes its first argument is when the entire loop is skipped, in which condition, according to (ii), $b_{1exit}$ equals $a$; in any other condition, $\phi_{L'}$ must take one of the other arguments, the value of which at the exit of the loop, according to (i), must equal $b_{1exit}$. This formula is useful for removing loop exit $\phi$ functions in the application of $\phi$-calculus as shown later. 

\subsubsection{Corollaries}
\label{sec:corollary}

At the bottom of Figure~\ref{fig:formulae} are some of the corollaries attained from the fundamental formulae. They provide facilities for transforming and simplifying $\phi$ expressions. 


The corollaries are in two groups. The first group consists of C1 to C4. These corollaries offer conveniences for symbolic differentiation, optimizations, and code generations, reducing computations and code size. Corollary C1 can be easily derived from the {\em distributed formula} through currying. Corollary C2 follows corollary C1 when we substitute $f$ with partial derivative.
Corollary C3 follows corollary C1 when we substitute $f$ with the $\phi$ function. Corollary C4 is attained when we apply C1 and then the {\em identity formula} (F1). 

The other group consists of corollaries C5 to C9, which offer conveniences for transforming $\phi$ expressions into closed forms for symbolic differentiation. 

Corollary C5 is proved as follows. 

\begin{proof}
Because of F4, we have the following relations:

\begin{align}
    d^{(1)} = f(\phi_L^{(1)}(p,d)) = f(p)\\
    d^{(2)} = f(\phi_L^{(2)}(p,d)) = f(d^{(1)}) = f(f(p)) = f^{[2]}(p)\\
    d^{(3)} = f(\phi_L^{(3)}(p,d)) = f(d^{(2)}) = f(f^{[2]}(p)) = f^{[3]}(p)\\
    ...\\
    d^{(n)} = f(\phi_L^{(n-1)}(p,d)) = f(d^{(n-1)}) = f(f^{[n-1]}(p)) = f^{[n]}(p)
\end{align}

Because of the definition of SSA, after the loop entry function $\phi_L$ in the final iteration of loop $L$, there shall be no other assignment to $d$ before the exit of the loop. Hence, $d_{exit(L)}=d^{(n)} = f^{[n]}(p)$. 
\end{proof}

Corollaries C6 to C9 are variants of C5 with function $f$ instantiated in several forms. They can be proved in a way similar to C5. 

\subsection{Examples}
\label{sec:phiExamples}

We now use several examples to show how $\phi$-calculus helps symbolic differentiation. We start with two simple ones and end with a more complicated case with nested loops, breaks, if-else, and arrays. \\

{\bf (I) If-Else Example.}  We first look at the example in Figure~\ref{fig:ifExample}. The $\phi$ functions resolve the difficulty for getting a closed form for the code. With the code in SSA, the derivation of the closed form for the code can simply ignore the conditional statements. What it needs to do is only to apply simple substitution of names with corresponding expressions based on the data flow. Figure~\ref{fig:ifExample}(d) shows the closed form obtained from the SSA form in Figure~\ref{fig:ifExample}(c). We add subscripts to the $\phi$ functions to help tell different $\phi$ functions apart. 

According to corollary C2, we can apply derivation on $x$ on the closed form and distribute the operation to the arguments of the $\phi$ functions. The result is shown in Figure~\ref{fig:ifExample}(e). The underlines indicate two simplification opportunities. (i) The first underlined expression, $(p+qx)*0.5*q*(p+qx)^{-0.5}$, can be easily simplified by symbolic engines into $0.5q(p+qx)^{0.5}$. (ii) The second simplification opportunity appears after the {\em distributive formula} (F2) is applied such that the final term ($1+e^{p+qx}$) in the expression in Figure~\ref{fig:ifExample} (e) is distributed into the $\phi$ functions. That term cancels the denominator of the second underlined expression. Figure~\ref{fig:ifExample} (f) shows the result after symbolic simplication. It is worth noting that such optimization opportunities appear because of $\phi$-calculus: They are both about interactions of the codelets across the boundaries of conditional branches, and hence would need the involved computations to be treated together. If each straight-line section of the codelet is symbolically differentiated individually as shown by the segments of interest (SOIs) in Figure~\ref{fig:ifExample}, those simplifications cannot get exposed. From Figure~\ref{fig:ifExample}(f), code can then be generated with the $\phi$ functions materialized with conditional statements that check the corresponding branching decisions recorded during the primal computation. 


{\bf (II) Simple Loop Example.}  Figure~\ref{fig:loopExample} (d) shows how $\phi$-calculus helps produce a closed form for the loop in Figure~\ref{fig:loopExample} (a). With $\phi$-calculus notations, the loop is symbolically represented in Figure~\ref{fig:loopExample} (c), on which, corollary C6 removes the loop notation and the loop entry $\phi$, and produces the closed form of $s_3$ at the exit of the loop: $a+k\times x$ ($k$ is the loop trip count). The application of {\em loop exit formula} F5 to the expression of $s_4$ removes the loop exit $\phi$ function, producing the simple expression $a+k\times x$. Symbolic differentiation can then be applied easily. For illustration purpose, this loop is made simple and the derivation of the closed form may resemble the recognition of induction variables in loop parallelizations~\cite{Tu+:ICS1995}. The next example gives a more thorough demonstration of the power of $\phi$ calculus.

\begin{figure}
    \centering
    \includegraphics[width=.7\textwidth]{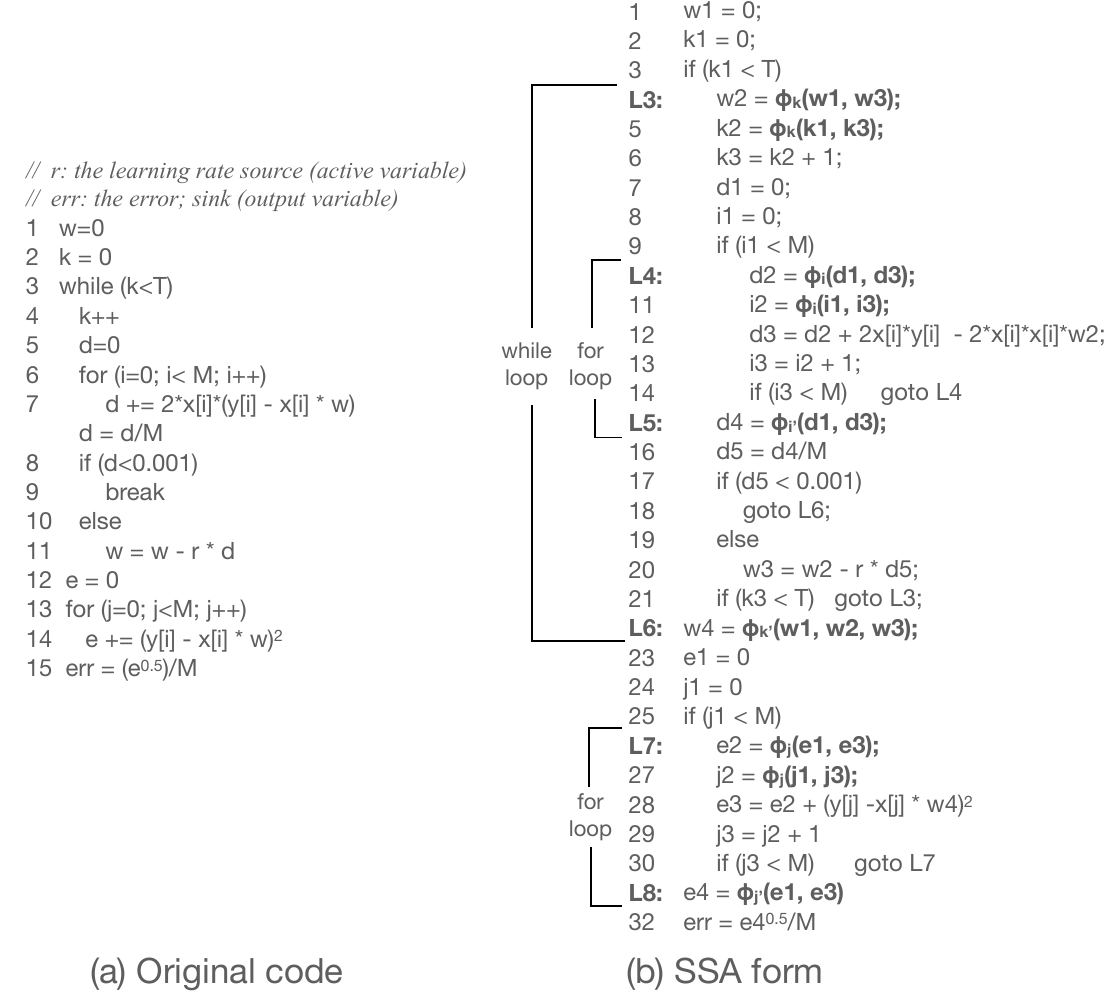}\\[4mm]
    \includegraphics[width=.98\textwidth]{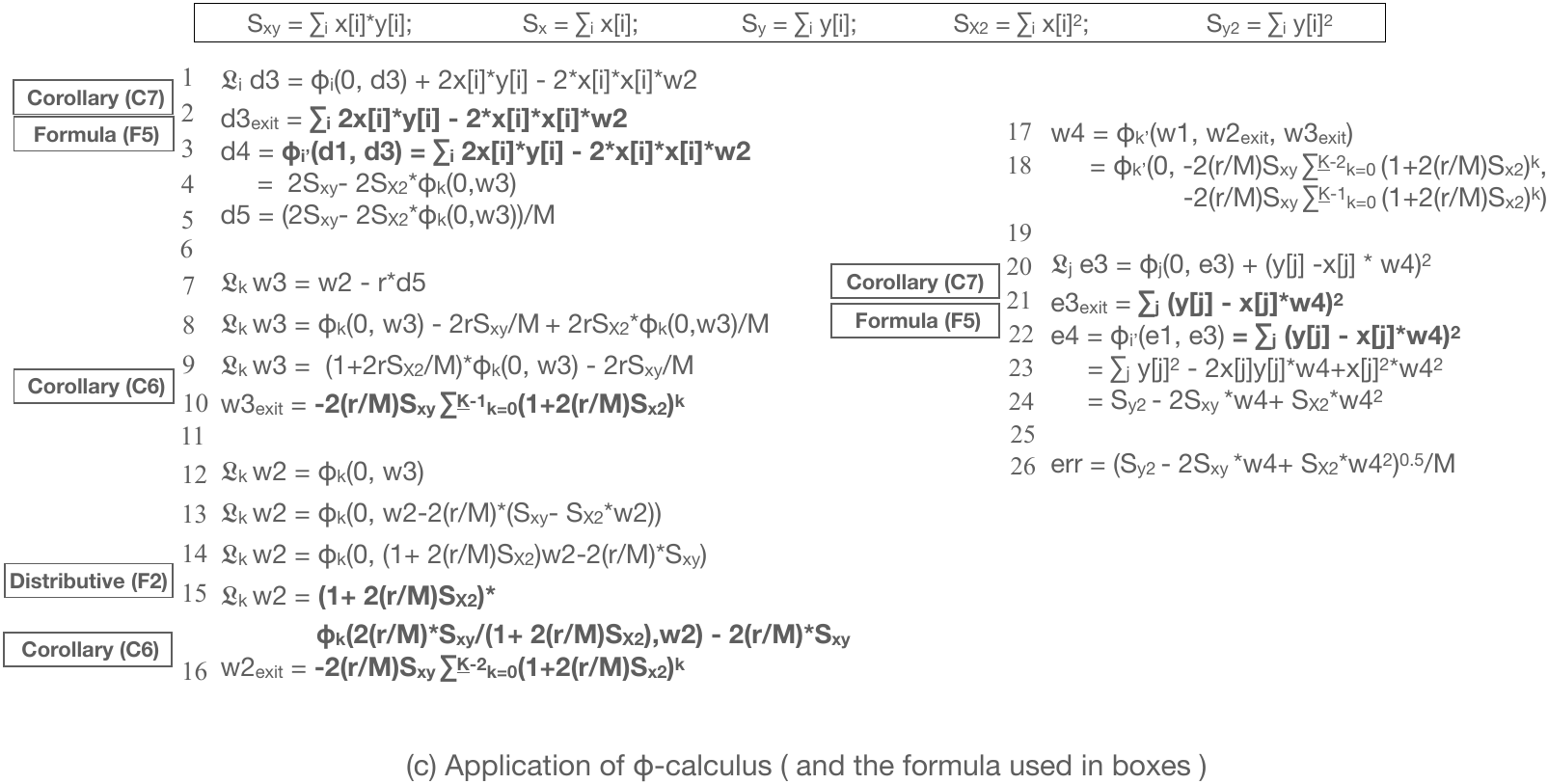}
    \caption{Application of $\phi$-calculus on hyperparameter optimizations, showing the treatment of loops and branches.} 
    \label{fig:bgdExample}
\end{figure}

{\bf (III) Complex Example (BGDHyperOpt).}  Figure~\ref{fig:bgdExample} shows a more complex example. The code in Figure~\ref{fig:bgdExample} (a) implements the use of batch gradient descent to determine the linear model on a dataset ($x$ for inputs, $y$ for response). The differentiation of interest is $d(err)/dr$, where $r$ is the learning rate; this gradient can be used in finding out the best learning rate---a so-called {\em meta learning} problem that optimizes hyperparameters of a machine learning process. 

The code consists of a \texttt{for} loop nested within a \texttt{while} loop; the \texttt{while} loop has a \texttt{break} in an \texttt{if-else} statement; there is another \texttt{for} loop following the \texttt{while} loop. To our best knowledge, no prior work can compute $d(err)/dr$ symbolically due to the control flow complexities.

Figure~\ref{fig:bgdExample}(b) shows the SSA form of the codelet. It includes nine $\phi$ functions; one of the loop exit $\phi$ functions ($\phi_{k'}$) has three arguments because of the \texttt{break} statement in the \texttt{while} loop. 

Figure~\ref{fig:bgdExample}(c) shows the application of $\phi$-calculus with the text boxes indicating the formulae or corollaries used at the important steps. We explain the process as follows. 

\textit{Lines 1-5:} Corollary C7 helps attain the closed-form expression of the value of $d3$ at the exit of the inner \texttt{for} loop; Formula F5 then resolves the $\phi_i'$ function and leads to the closed-form expression of $d5$. 

\textit{Lines 7-10:} This part tries to get the closed-form expression for the third argument ($w3$) of the $\phi_{k'}$ function on Line L6 in Figure~\ref{fig:bgdExample}(b). The part starts with a series of substitutions based on the results from Lines 1-5 in Figure~\ref{fig:bgdExample}(c), and then uses corollary C6 to get the closed-form expression of the value that $w3$ has at the normal (rather than via \texttt{break}) exit of the while loop. 

\textit{Lines 12-16:} This part tries to get the closed-form expression for the second argument ($w2$) of $\phi_{k'}$. It starts with substitutions with the results obtained already. It then uses the {\em distributive formula} F2 to transform the $\phi_k$ function on Line 14 in Figure~\ref{fig:bgdExample}(c) to a form matching the lhs form in corollary C6. The transformation is to factor out the terms in the second argument of $\phi_k$ such that the second arguement turns into pure $w2$ as shown on Line 15. Then corollary C6 can be applied, resolving the loop notation and also the $\phi_k$ function and producing the closed-form expression of $w2$ at the exit of the while loop as shown on Line 16 in Figure~\ref{fig:bgdExample}(c) (\underline{K} stands for the trip count of the while loop). 

\textit{Lines 17-18:} simple substitutions of the three arguments in $\phi_k'$.

\textit{Lines 20-24:} This part tries to get the closed-form expression for $e4$. It first applies corollary C7 to the expression on Line 20 in Figure~\ref{fig:bgdExample}(c) to resolve $\phi_j$ and the loop notation, producing the closed-form expression for $e3_{exit}$ on Line 21. It then applies formula F5 to resolve $\phi_j'$ on Line 22, producing the closed-form expression on Line 24 for $e4$. 

\textit{Line 26:} A simple substitution with the results produced so far gives the closed-form expression of the final variable $err$. (For the sake of readability, we leave out the substitution of $w4$.) Symbolic differentiation can then be applied to $err$ on $r$. 

\vspace{.1in}
This part has demonstrated the applications of $\phi$-calculus to several concrete examples. We will present the general use of $\phi$-calculus in the overall algorithm of coarsening in the next section.

\section{SOI Identification}
\label{sec:segmentation}

With $\phi$-calculus removing the barriers of control flow for coarsening, a {\em segment of interest (SOI)}---that is, the segment of code for symbolic differentiation---of a program can be much larger than a basic block. A larger SOI often offers more opportunities for optimizations, but it is not always better due to a tradeoff caused by two factors. 

The first is "expression swell". As aforementioned, "expression swell" refers to the phenomenon that the derivative often has a much larger (in the worst case, quadratically larger) representation than the original function has~\cite{Corliss:Swell1988}. As a result, a very long expression can cause large memory usage and long running time of symbolic engines. 

The second factor is the tension between coarsening and computation reuse. In coarsened differentiation, without a careful design, some computation reuse opportunities could get lost due to computation reordering caused by coarsening transformations, a phenomenon we call {\em reuse deprivation}. 


{\em Deprivation Example.} The impact of reuse deprivation can be seen on $x_{t+2,0}$ and $x_{t+2,1}$ in the \texttt{CartPole} example in Figure~\ref{fig:cartpole}. As Figure~\ref{fig:cartpole}(c) shows, they are both computed from $x_{t+1,1}$. So potentially, if $d(x_{t+2,0}/d(a_t)$ has been computed, $d(x_{t+1,1})/d(a_t)$ could be known and could be reused in computing $d(x_{t+2,1}/d(a_t)$.  Coarsening the computations from $a_t$ and $X_t$ (i.e., [$x_{t,0}, x_{t,1}, x_{t,2}, x_{t,3}$]) to $x_{t+2,0}$, however, deprives that reuse opportunity. The coarsening result has been shown in Equation~\ref{eq:xt2}, in which the holders of intermediate results, such as $x_{t+1,1}$, disappear. The differentiation over $a_t$ is shown in Equation~\ref{eq:dxt2}, which has no $d(x_{t+1,1})/d(a_t)$ or the derivatives of any other intermediate variables over $a_t$. As a result, when we need to compute the derivative of $x_{t+2,1}$ over $a_t$, we cannot reuse those intermediate derivatives.

The example illustrates a tension between reuse and coarsening. The larger is the coarsening granularity, the more opportunities there are for the enabled symbolic differentiation and optimization to take effect, but at the same time, it could incur deprivation of computation reuse opportunities. 


What adds subtly to the relation is that reuse deprivation does not always lead to fewer reuse opportunities. Some reuse deprivations transform the reuse opportunities to another form. For instance, suppose that we have a way to get a closed form for the entire computation from $X_t$ to $L$. All reuses, including $x_{t+1,1}$ for $x_{t+2,0}$ and $x_{t+2,1}$, turn into explicit sub-expressions in the closed form of $L$. There can hence be reuse opportunities exposed between them in the differentiation of the closed-form expression. The condition for such a transformation of reuse to occur is that the closed-form expression must subsume both parties that contain the reusable computations. 

A single solution, reuse-aware SOI identification, addresses both factors. {\em Reuse-aware SOI identification} refers to an algorithm that solves the following optimization problem:

\begin{definition}\label{def:reuseAwareSeg}
{\bf Optimal SOI segmentation problem:} Let $G$ be a series of computations, $l$ be the upper limit of the allowed sizes of an SOI, $P$ be the set of valid partitions of $G$, that is, for any partition $S$ in $P$, no element in $S$ is larger than $l$. The problem is to find the optimal partition $S^*\in P$ \rev{such that the total running time is minimized, that is,

\[
\forall Q\in P, ad(S^*)+\sum_{s\in S^*}compute(\mathit{dif(s)})\le ad(Q)+\sum_{q\in Q} compute(\mathit{dif(q)}),\] }

\noindent where $compute(\mathit{dif(x)})$ is the amount of computation involved in running the symbolically differentiated code segment for code $x$, and $ad(X)$ is the cost of the remaining AD differentiation of $X$ after symbolic differentiation.
\end{definition}

The upper bound of SOI size $l$ in the problem description ensures that the symbolic engine works well even in the presence of the "expression swell" effects. The problem description indicates three factors relevant to SOI definitions.

1) The cost $ad(X)$. This cost is incurred at the boundaries of SOIs. Symbolic differentiation of an SOI computes only the derivatives of the active output variables of this SOI on the active input variables of this SOI. These derivatives have to be connected into a chain by AD to compute the derivatives of the ultimate active output variables on the ultimate active input variables. As a result, the more SOIs there are, the more AD overhead is there, and the larger is $ad(X)$. In the extreme case where each operation is an SOI, $ad(X)$ would equal to the cost taken by the default AD without coarsening. So this factor calls for larger SOIs.

2) Computation simplifications. The cost $\sum_{x\in X}compute(\mathit{dif(x)})$ is smaller if more computations are simplified. As two cancellable computations falling into two separate SOIs are not going to get cancelled by the symbolic engine, maximization of simplified computations also calls for larger SOIs. 

3) Computation reuse. Maximizing computation reuse helps reduce the cost $\sum_{x\in X}compute(\mathit{dif(x)})$ as well. Unlike computation simplification, computation reuse exists both within and across SOIs. Exploiting reuse within an SOI happens in the default symbolic optimization and code optimization (e.g., common subexpression elimination (CSE)~\cite{dragonBook}). Reuse across SOIs is the natural result of the chain rules of differentiation, as shown by the potential reuse of $d(x_{t+1,1})/d(a_t)$ in computing $d(x_{t+2,0})/d(a_t)$ and $d(x_{t+2,1})/d(a_t)$ in Figure~\ref{fig:cartpole}. In general, if $y$ is an active output variable of both $SOI_a$ and $SOI_b$ and it is also an active input variable of $SOI_c$, then the derivative of the ultimate active output $z$ on $y$ ($dz/dy$) can be used in computing the derivatives of $z$ on the active inputs of both $SOI_a$ and $SOI_b$. Besides saving computations, reuses are also helpful for mitigating the ``expression swell" problem as they split a long expression into shorter ones as noted in some previous work~\cite{wang2019demystifying,laue2020equivalence}. Because expanding SOIs could lose inter-SOI reuses as the {\em deprivation example} has shown, this factor suggests that simply maximizing SOIs to the upper limit $l$ cannot always give the best SOIs. 

Finding the optimal SOI segmentation is difficult. \rev{For an SSA representation with $N$ instructions, assuming every instruction can be put into an SOI, the number of possible partitions is between $l^{N/l}$ and $l^N$, where, $l$ is the upper limit of the allowed size of an SOI. The reason is that the number of SOIs is between $N/l$ and $N$, while the size of an SOI has $l$ possibilities in the lower-bound case and up to $l$ possibilities in the upper-bound case.} Besides the exponential space, it would require detailed performance and overhead modeling, which is hard to be precise at static compile time. 

\begin{figure}
\centering
\includegraphics[width=.9\textwidth]{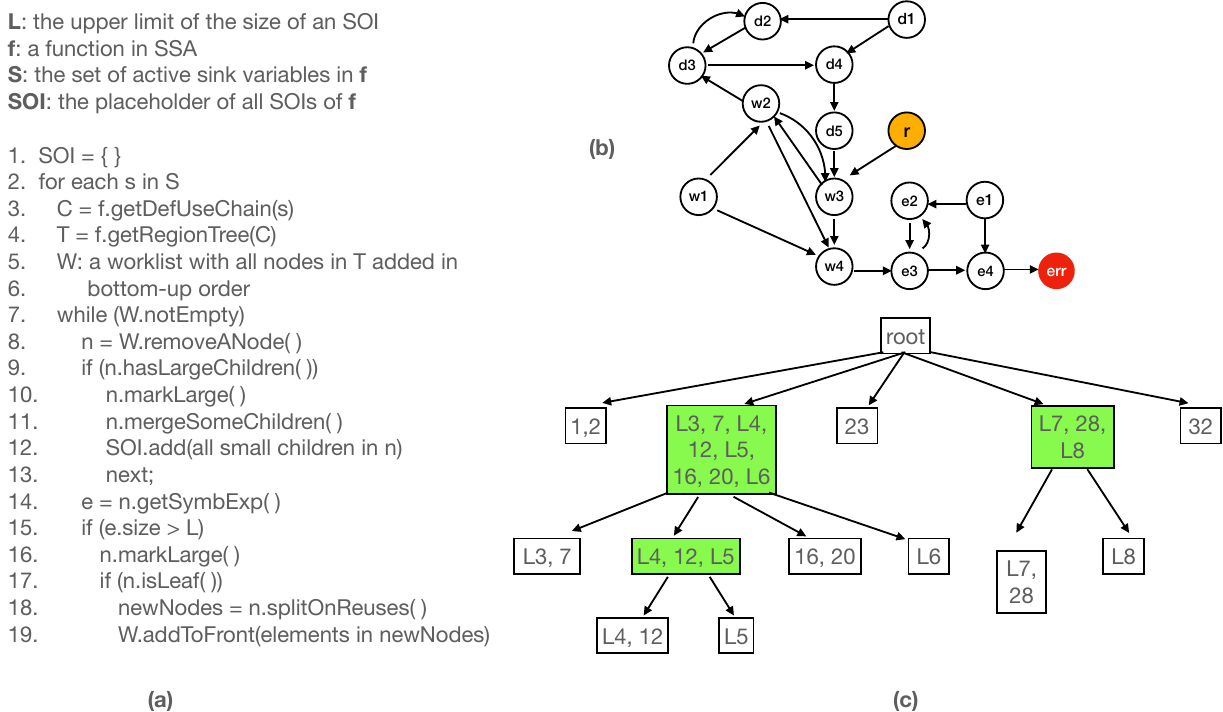}
\caption{(a) Algorithm of reuse-aware identification of SOIs for coarsening. (b) The def-use chain of {\em err} in the example shown in Figure~\ref{fig:bgdExample}(b) with $r$ as the active input; inactive inputs ($x$, $y$, $M$) are omitted. (c) The {\em def-use region tree} for $err$ in Figure~\ref{fig:bgdExample}(b); each element in a box is a line number in Figure~\ref{fig:bgdExample}(b); solid (green) boxes stand for loops.}\label{fig:soiAlg}
\end{figure}

It is however important for a viable solution to take all these factors into consideration. 
Figure~\ref{fig:soiAlg}(a) outlines our designed algorithm. For a given function $f$, for each of its active variable $s$ that outlives $f$, the algorithm gets its def-use chain, which captures all the definitions in $f$ that lead to the value of $s$. It then builds a {\em def-use region tree} out of all the definitions on the def-use chain. {\em Def-use region tree} is a data structure inspired by the classic {\em code region hierarchy} in compilers~\cite{dragonBook}. Traditionally, a code region is defined as a collection of nodes $N$ and edges $E$ such that (i) a header node $h$ in $N$ dominates all other nodes in the collection; (ii) if $p$ is in $N$, then m must be in $N$ if $m$ reaches $p$ without going through $h$; (iii) $E$ includes all edges between nodes in $N$, except for those that enter $h$.
 In the code region hierarchy of a program, each code region is represented by a node in the hierarchy subsumed under the nodes that represent its enclosing code regions. {\em Def-use region tree} has two major differences from {\em code region hierarchy}: (i) only relevant variable definitions are considered; (ii) every loop-exit $\phi$ function is put as part of the region of the associated loop. The second property is for convenience in the derivation of symbolic expressions for loops. As an example, Figure~\ref{fig:soiAlg}(b) shows the def-use region tree of all the definitions on the def-use chain of $err$ in Figure~\ref{fig:bgdExample}(b); each element in the boxes is a line number in Figure~\ref{fig:bgdExample}(b). The three solid (green) boxes are the three loops. The three loop-exit $\phi$ functions (Lines L5, L6, L8 in Figure~\ref{fig:bgdExample}(b)) are put together with the three loops respectively. 

The SOI identification algorithm then traverses the region tree in a bottom-up order, as outlined in Figure~\ref{fig:soiAlg}(a). For each node, if it has no large child node (i.e., exceeding the SOI size limit), its symbolic expression is derived through the $\phi$-calculus. If the size of the derived expression exceeds the SOI size limit, that node is marked as a large node, and if it is a leaf node, it is split into two smaller nodes; the splitting point is chosen to be the variable that is contained in that node and has the largest number of references (and hence reuses) in $f$. The two new nodes created by the split are added to the front of \texttt{worklist}. If the current node has large children, there is no need to go up further in the {\em def-use region tree} as the upper nodes can only become even larger. In that case, the algorithm examines the immediate children of this node, merge consecutive small children nodes (up to the SOI size limit); after that, it puts each of the children nodes smaller than the limit as an SOI. The algorithm continues until {\em worklist} becomes empty. The size limit $L$ can be empirically selected based on the machine and the symbolic engine.

The algorithm follows the principle of maximizing the size of SOIs within the SOI size limit while respecting reuses when it is necessary to split a def-use chain into multiple SOIs. 


\section{Implementation}

We implement the coarsening optimization on an in-house AD tool named DiffKt. The tool was developed for Kotlin, a cross-platform, statically typed, general-purpose programming language with type inference. The tool itself is also written in Kotlin. We choose it as the basis mainly because of its availability and the statically typed nature of Kotlin which offers conveniences for static code analysis and transformations. But as a general optimization technique, coarsening can be potentially applied to many other AD tools; for some (e.g., Python AD tools), it may need to be done dynamically. 

DiffKt was developed and optimized by 10+ engineers in industry in over a year. It supports both CPU and GPU, with high-performance native math/DNN libraries for Tensor computations. The tool is planned to open source in the near future. A systematic benchmarking of the tool over other AD tools is yet to be done, but preliminary measurements show that it outperforms PyTorch AD~\cite{pytorchDiff} by over 10$\times$ on scalar-intensive cases (e.g., \texttt{HookeanSpring} in Section~\ref{sec:evaluation}), and achieves comparable speeds on common deep learning models where pre-existing libraries are called for gradients calculations of the standard DNN layers under the hood of both of them. (As PyTorch AD and the Kotlin AD tool are in different programming languages, the comparison is only to give readers a sense about the industrial quality of the default tool.) 

Similar to many other AD tools (e.g., PyTorch~\cite{pytorchDiff}, JAX~\cite{jax2018github}), DiffKt is a library-based implementation, enabling AD through operator overloading via a generic class \texttt{Tensor}. Implemented in 250K lines of code, it supports backward AD and includes a Tensor typing module as well. 

\rev{As with other Automatic Differentiation (AD) tools (e.g., Zygote~\cite{Innes:MLSys2020}), DiffKt also allows the use of adjoints for custom differentiation. For a given expression $e$, if a custom differentiation of $e$ is provided, the AD process will automatically invoke it rather than conduct the default operation-by-operation differentiation. The differentiations of the SOIs generated by coarsening are integrated into the original program as custom adjoints. An example is shown in Figure~\ref{fig:adjoint}. The coarsening results form the content of the adjoint, which is associated with the coarsened primal expression through the call "setIntermediateAdjoints". If coarsening produces code for differentiating the entire primal code, the compiler simply replaces the calls of the corresponding \texttt{backward} function with the generated code; if the compiler in addition determines that the primal results are used in the program only for getting the derivatives, the compiler removes the invocations of the primal code.

\begin{figure}
    \centering
    \includegraphics[width=0.95\textwidth]{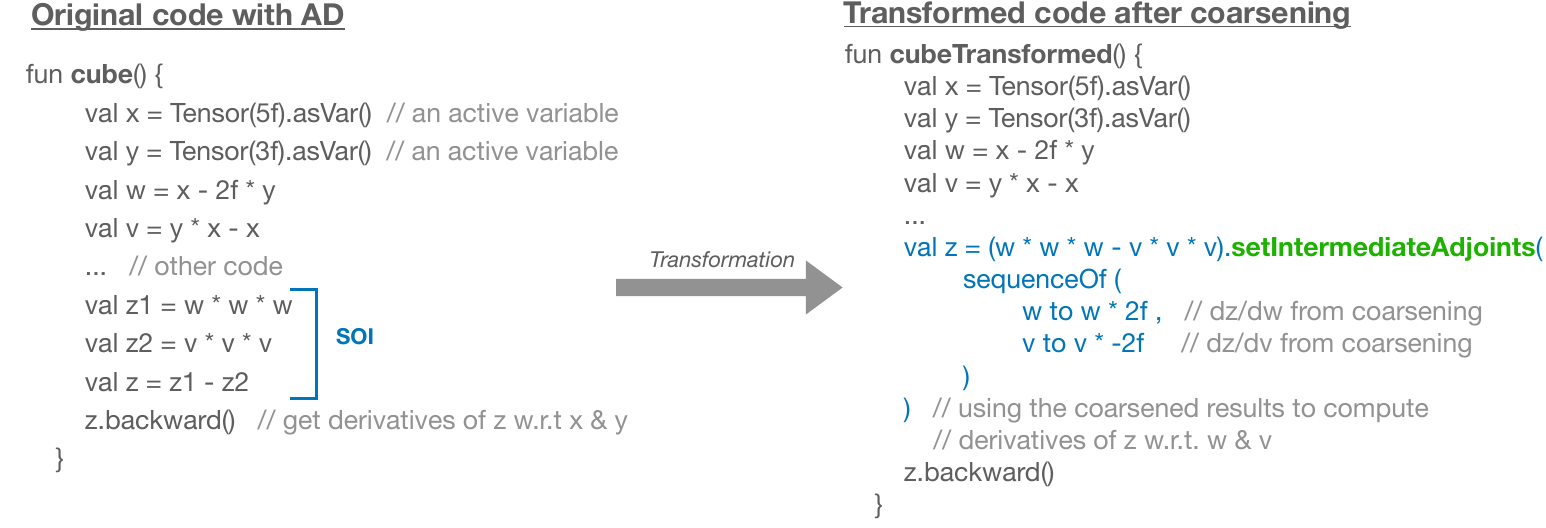}
    \caption{\rev{An example showing how the coarsened results are integrated into the original program. The original primal code (left) is transformed to a form (right) such that the calculations of $z$ in the SOI are put into one expression, and an adjoint is appended to that expression, which, at runtime, makes the AD take its content as the shortcuts for calculating the derivatives of $z$ on the two active variables of the SOI, $w$ and $v$. The results are used by the default AD in differentiating the rest of the code (i.e., from $w$ and $v$ to $x$ and $y$.}}
    \label{fig:adjoint}
\end{figure}
}

Our implementation of coarsening on DiffKt is based on a Kotlin compiler. \rev{Our experiments focus on first-order backward AD, but it is worth noting that coarsening, as a way to offer shortcuts in AD, can in principle help higher-order differentiation and forward or mixed-direction AD as well.} In addition to the $\phi$-calculus and the SOI identification as presented earlier, our implementation also includes loop unrolling and the use of the primal computation results in the generated adjoint functions when possible. As a side benefit of coarsening, our implementation of coarsening optimizes the primal in addition to the differentiation: After getting the closed form of the primal computation, it applies symbolic optimization to the primal and regenerates the code; \rev{an example is shown in the BGDHyperOpt benchmark in Section~\ref{sec:results}.}  In the case where only differentiation is needed and the entire primal can be symbolically differentiated, the optimizer removes the primal from the program when possible---an optimization elusive to existing AD. In the current implementation, for proof of concept, we use a symbolic engine extended from Sympy~\cite{sympy}, an open-source symbolic manipulation tool. 

\rev{Like many other compiler-based optimizations that change the order of computations, coarsening could affect the numerical precision. A convenience offered by coarsening is that measures to avoid numerical unstableness can be seamlessly integrated into the code generation in coarsening. The code generator is equipped with the patterns for dealing with common numerically unstable expressions. Before it generates the code for a symbolic expression, it examines it to identify the numerical unstable expressions through pattern matching, and generates the code corresponding to their numerically stable forms. For expression $log(1+e^{-x\beta})$, for example, as the code generator finds out that the expression matches one of the patterns in its unstable list, $log(1+e^n)$, it generates the code to discern the value of the exponent, as illustrated as follows (MAXEXP is set to 40 in our implementation): 

\begin{quote}
temp0 = -x$\beta$ \\
temp1 = (temp0 > MAXEXP)? MAXEXP : log(1+$e^{temp0}$)
\end{quote}

When the expression operates on tensors, the generated code uses masking functions (like \texttt{where} in PyTorch) for efficiency. A concrete example of numerically stable code generation in coarsening is the \texttt{HMC} benchmark detailed in Section~\ref{sec:results}.}

\section{Evaluation}
\label{sec:evaluation}

To evaluate the efficacy of the proposed techniques, we test coarsening on six applications in 18 total configurations on two different machines. {\em Backward AD is used.} The results show that the optimization improves the differentiation speed by 1.03-27$\times$, and the whole application execution speed by 1.08-11$\times$. 

\subsection{Methodology}

\begin{table}
\centering
\caption{Benchmarks and Configurations}\label{tab:benchmark}
\footnotesize
\begin{tabular}{|c|p{.8in}|p{1.5in}|c|p{1.5in}|}\hline
{\bf Name} & {\bf Domain} & {\bf Description} & \multicolumn{2}{c|}{\bf Configs} \\\hline
BGDHyperOpt & Meta- & Optimizing the learning rate &  1 & 200 data records \\ 
& Learning& of batch gradient descent  & 2 & 1000 data records \\ 
& & based linear regression & 3 & 2000 data records \\ \hline
Brachist. & Math. & Brachistochrone curve & 1 & 200 data points \\
& Physics &  calculation & 2 & 400 data points \\
& & & 3 & 1000 data points \\ \hline
CartPole & Deep  & Training a CartPole system &  1 & an update every 6 steps\\ 
& Reinforcement & & 2 & an update every 8 steps \\ 
& Learning & & 3 & an update every 10 steps \\ \hline
HMC & Statistic & Hamiltonian Monte Carlo & 1 & 100 one-dim records \\
& Sampling  &Sampling for logistic  & 2 & 1000 two-dim records \\
& for Prob. Prog. &regression  & 3 & 800 three-dim records \\ \hline
HookeanSpring & Physical & Simulating the dynamics of a & 1 & 10 vertices \\ 
& Simulation &Hookean Springs system  & 2 & 20 vertices \\ 
& & & 3 & 40 vertices \\ \hline
QWOP & Gaming  & An avatar learns walking & 1 & light-weight figure \\
&  &via motion optimization & 2 & medium-weight figure\\
&  & & 3 & heavy-weight figure \\\hline
\end{tabular}
\end{table}

\begin{table}
\centering
\footnotesize
\caption{Machines}\label{tab:machines}
\begin{tabular}{|c|p{4.5in}|}\hline
{\bf Machine} & {\bf Configuration} \\\hline
devServer & Intel(R) Xeon(R) Gold 6138 40-core CPU \@ 2.00GHz, 250GB, CentOS Stream 8, Kotlin 1.4.20-M1, Java HotSpot 64-Bit Server VM, Java 1.8.0\_192\\\hline
Macbook & MacBook Pro, 2.4GHz 8-core Intel Core i9, 32GB 2667MHz DDR4, MacOS Catalina (v. 10.15.7), Kotlin 1.4.20-M1, Graalvm 20.3.0, Java 11.0.9\\\hline
\end{tabular}
\end{table}



\paragraph{Benchmarks} There are no common benchmark suites designed for evaluating AD. We collected six applications from several domains where AD is important, and implemented them in Kotlin with the Kotlin AD library. Table~\ref{tab:benchmark} lists the set of benchmarks used in the experiments. These benchmarks come from several domains, from physical simulation to statistical sampling, deep reinforcement learning, gaming, and meta learning. They also show a range of code complexities, with BGDHyperOpt featuring control flow complexities as Figure~\ref{fig:bgdExample} has shown, Brachist. featuring a case where primal computation could be potentially removed, CartPole featuring a combination of matrix-based Deep Neural Network and scalar-based environment simulations, HMC featuring potential value overflow incurred by exponential computations, HookeanSpring featuring a sequence of regular vector operations, and QWOP featuring a long function with many small loops and if-else statements. For each benchmark, we include three configurations as listed in the right column of Table~\ref{tab:benchmark}, which will be explained later in the discussion of the results of each application. We repeat the performance measurements multiple times and report both the mean and standard deviation of the timing results. Kotlin runs on Java virtual machines. For both the baseline and the optimized versions, the JRE went through a warm-up phase before timing starts to get the stable performance.

\paragraph{Machines} Because in practical scenarios, those AD-based applications may run on both servers and personal computers/laptops, we have measured the performance of the applications on both kinds of machines. Table~\ref{tab:machines} provides the machine details.

\subsection{Results}
\label{sec:results}

In this part, we first present an overview of the performance, and then provide detailed discussions on each benchmark.

Table~\ref{tab:results} reports the overall performance, where "baseline" represents executions of the default Kotlin AD tool and "opt" represents executions after coarsening is applied. The "Differentiation Time" column reports the time taken by differentiation in one iteration of each benchmark, while the "Overall Time" column reports the overall time of an iteration. We make two observations.

(1) Coarsening brings 1.03--27$\times$ speedups to the differentiation of the benchmarks, and 1.08-11$\times$ speedups to the overall execution. In most cases, the overall speedups are smaller than the differentiation speedups as there are some parts of the computation in the programs outside the part of the code targeted by the coarsening optimization (i.e., the part involved in differentiation). Exceptions are \texttt{Brachist.} and \texttt{HookeanSpring}; it is because in those two original programs, the only purpose of the primal computations are to let the AD to compute the gradients. Because coarsening generates code that can directly computes the gradients, the optimization removes the primal computations completely; the overall time is hence shortened even more than the time savings on the differentiation. (A side observation is that in all cases, the laptop runs faster than the server, probably due to its faster CPUs and the use of a more recent version of Java Runtime.)

(2) Coarsening is consistently beneficial; it saves the execution time across benchmarks, configurations, and machines. The main reasons for the time savings are three: (i) the savings of the operator overloading overhead of AD, which comes from object boxing and memory allocations; (ii) the simplifications of the computations thanks to the large-scoped symbolic differentiation and optimizations; (iii) the removal of unnecessary primal computations. We next elaborate these benefits through in-depth examinations of each of the benchmarks.


\begin{table}
\footnotesize
\centering
\caption{Experimental Results: Time per iteration and Speedups}\label{tab:results}
\begin{tabular}{|c|c|c||c|c|c||c|c|c|}\hline
 & & & \multicolumn{3}{c||}{Differentiation Time(ms)} & \multicolumn{3}{c|}{Overall Time(ms)}\\\cline{4-9}
Machine & Benchmark & Config & baseline & opt & speedup & baseline & opt & speedup \\\hline
devServer & BGDHyperOpt & 1 & 5.44$\pm$  6\% & 0.20$\pm$  1\% & 27.44X & 10.44$\pm$  5\% & 1.22$\pm$  2\% & 8.52X \\& & 2 & 5.22$\pm$  8\% & 0.20$\pm$  2\% & 26.18X & 10.12$\pm$  7\% & 1.21$\pm$  2\% & 8.37X \\& & 3 & 5.37$\pm$  6\% & 0.20$\pm$  2\% & 26.82X & 10.36$\pm$  6\% & 1.21$\pm$  2\% & 8.56X \\ \cline{2-9}
& Branchist. & 1 & 0.04$\pm$  3\% & 0.04$\pm$  7\% & 1.10X & 0.09$\pm$  3\% & 0.04$\pm$  7\% & 2.51X \\& & 2 & 0.19$\pm$ 36\% & 0.15$\pm$  1\% & 1.22X & 0.32$\pm$ 34\% & 0.15$\pm$  1\% & 2.08X \\& & 3 & 0.55$\pm$  2\% & 0.39$\pm$  4\% & 1.41X & 0.69$\pm$  2\% & 0.39$\pm$  4\% & 1.79X \\ \cline{2-9}
& CartPole & 1 & 52.15$\pm$  4\% & 46.86$\pm$  1\% & 1.11X & 55.06$\pm$  4\% & 49.38$\pm$  1\% & 1.12X \\& & 2 & 15.69$\pm$  1\% & 14.51$\pm$  1\% & 1.08X & 16.99$\pm$  1\% & 15.65$\pm$  1\% & 1.09X \\& & 3 & 15.77$\pm$  2\% & 14.06$\pm$  2\% & 1.12X & 17.07$\pm$  2\% & 16.24$\pm$  2\% & 1.05X \\ \cline{2-9}
& HMC & 1 & 2.89$\pm$  5\% & 0.67$\pm$  2\% & 4.29X & 3.26$\pm$  5\% & 0.94$\pm$  3\% & 3.46X \\& & 2 & 4.55$\pm$  1\% & 1.59$\pm$  4\% & 2.86X & 5.17$\pm$  1\% & 2.11$\pm$  4\% & 2.45X \\& & 3 & 5.58$\pm$  7\% & 2.19$\pm$  5\% & 2.54X & 6.35$\pm$  7\% & 2.79$\pm$  5\% & 2.27X \\ \cline{2-9}
& HookeanSpring & 1 & 0.10$\pm$  4\% & 0.01$\pm$  2\% & 6.62X & 0.16$\pm$  5\% & 0.01$\pm$  2\% & 11.02X \\& & 2 & 0.11$\pm$  8\% & 0.03$\pm$  2\% & 4.09X & 0.18$\pm$  8\% & 0.03$\pm$  2\% & 6.72X \\& & 3 & 0.23$\pm$ 12\% & 0.05$\pm$ 13\% & 4.52X & 0.38$\pm$ 10\% & 0.05$\pm$ 13\% & 7.53X \\ \cline{2-9}
& QWOP & 1 & 1.97$\pm$  6\% & 1.46$\pm$  4\% & 1.35X & 3.89$\pm$  4\% & 2.76$\pm$  3\% & 1.41X \\& & 2 & 25.71$\pm$  5\% & 17.81$\pm$  6\% & 1.44X & 44.50$\pm$  3\% & 29.61$\pm$  4\% & 1.50X \\& & 3 & 32.17$\pm$  5\% & 21.20$\pm$  6\% & 1.52X & 56.86$\pm$  3\% & 36.37$\pm$  4\% & 1.56X \\ \hline
macBook & BGDHyperOpt & 1 & 3.59$\pm$  4\% & 0.14$\pm$  2\% & 25.15X & 7.23$\pm$  4\% & 0.86$\pm$  2\% & 8.41X \\& & 2 & 3.59$\pm$  3\% & 0.15$\pm$  6\% & 23.43X & 7.31$\pm$  3\% & 0.91$\pm$  4\% & 8.06X \\& & 3 & 3.66$\pm$  4\% & 0.15$\pm$  3\% & 24.23X & 7.41$\pm$  3\% & 0.91$\pm$  4\% & 8.18X \\ \cline{2-9}
& Branchist. & 1 & 0.04$\pm$ 18\% & 0.04$\pm$ 12\% & 1.03X & 0.10$\pm$ 14\% & 0.04$\pm$ 12\% & 2.38X \\& & 2 & 0.13$\pm$  4\% & 0.12$\pm$  2\% & 1.08X & 0.23$\pm$  4\% & 0.12$\pm$  2\% & 1.91X \\& & 3 & 0.44$\pm$  2\% & 0.31$\pm$  5\% & 1.42X & 0.55$\pm$  2\% & 0.31$\pm$  5\% & 1.80X \\ \cline{2-9}
& CartPole & 1 & 29.46$\pm$  3\% & 26.42$\pm$  2\% & 1.12X & 31.16$\pm$  3\% & 27.92$\pm$  2\% & 1.12X \\& & 2 & 9.43$\pm$  2\% & 8.76$\pm$  0\% & 1.08X & 11.89$\pm$  2\% & 11.03$\pm$  1\% & 1.08X \\& & 3 & 9.06$\pm$  0\% & 8.35$\pm$  1\% & 1.09X & 12.14$\pm$  1\% & 11.25$\pm$  0\% & 1.08X \\ \cline{2-9}
& HMC & 1 & 2.60$\pm$  1\% & 0.59$\pm$  2\% & 4.42X & 2.90$\pm$  1\% & 0.82$\pm$  2\% & 3.56X \\& & 2 & 3.88$\pm$  1\% & 1.24$\pm$  1\% & 3.13X & 4.37$\pm$  1\% & 1.64$\pm$  1\% & 2.67X \\& & 3 & 4.30$\pm$  1\% & 1.61$\pm$  1\% & 2.67X & 4.85$\pm$  1\% & 2.06$\pm$  0\% & 2.35X \\ \cline{2-9}
& HookeanSpring & 1 & 0.08$\pm$  8\% & 0.01$\pm$  7\% & 5.72X & 0.14$\pm$  8\% & 0.01$\pm$  7\% & 9.84X \\& & 2 & 0.09$\pm$ 11\% & 0.02$\pm$  2\% & 4.17X & 0.15$\pm$  8\% & 0.02$\pm$  2\% & 7.13X \\& & 3 & 0.10$\pm$  2\% & 0.04$\pm$  4\% & 2.46X & 0.17$\pm$  2\% & 0.04$\pm$  4\% & 4.09X \\ \cline{2-9}
& QWOP & 1 & 1.57$\pm$  3\% & 1.16$\pm$  2\% & 1.35X & 3.60$\pm$  4\% & 2.45$\pm$  2\% & 1.47X \\& & 2 & 20.53$\pm$  2\% & 14.70$\pm$  2\% & 1.40X & 40.43$\pm$  2\% & 26.46$\pm$  2\% & 1.53X \\& & 3 & 24.69$\pm$  2\% & 16.81$\pm$  1\% & 1.47X & 50.01$\pm$  2\% & 31.89$\pm$  1\% & 1.57X \\ \hline

\end{tabular}
\end{table}



\noindent{\bf BGDHyperOpt.} BGDHyperOpt is a meta-learning program and has been introduced in Example III in Section~\ref{sec:phiExamples} and Figure~\ref{fig:bgdExample}. Meta-learning entails inspecting and optimizing a machine learning process, which has recently drawing lots of interest. This program tries to optimize learning rates through gradient descent. The three configurations correspond to three different sets of inputs.

\rev{The SOI in this program is the entire function that computes the learning error as shown in Figure~\ref{fig:bgdExample}(a). There are nine $\phi$ functions in the SOI.} Despite the control complexities, coarsening is able to get the closed-form expression for the entire gradient computation and apply symbolic differentiation on it, giving more than 23$\times$ differentiation speedups in all cases. The primal cannot be removed because the generated differentiation code must use the trip-counts of the \texttt{while} loop in the primal. As a result, the overall speedup is about 8$\times$. Meanwhile, coarsening saves over 70\% of memory allocations because many of the data allocations in the default AD are for holding intermediate data objects which are no longer needed after coarsening. 

We did an ablation study to examine the benefits from the $\phi$-calculus. In the study, we apply coarsening without $\phi$-calculus; symbolic differentiation is hence applied to only the inner-most loop (which is written as a single Tensor statement) and the code after the \texttt{while} loop. The differentiation speedups drop from 23-27$\times$ to 8-9$\times$:

    \begin{quote}
    \small
    \begin{tabular}{c|c|c|c|c|c|c}\hline
         &  \multicolumn{3}{c|}{devServer} & \multicolumn{3}{c}{macBook} \\\cline{2-7}
    Config   & 1 & 2 & 3 & 1 & 2 & 3 \\\hline
    Speedup(X) w/o $\phi$-calculus &     8.46	& 7.80	& 8.08	& 9.24	& 9.34	& 9.39\\
    Speedup(X) w/ $\phi$-calculus & 27.44	& 26.18	& 26.82	& 25.15	& 23.43	& 24.23\\\hline
    \end{tabular}
    \end{quote}
It can be seen that the larger scope of optimizations enabled by $\phi$-calculus boosts the speedups by a factor of three. Its effects are multi-fold: (i) It allows a complete removal of the boxing overhead of the Tensor data structure from the differentiation process, whereas without $\phi$-calculus, as only part of the differentiation is symbolically done, Tensors have to be used so that the operator overloading can still work, which is what the remaining part of the AD depends on. (ii) It exposes large-scoped loop-invariant calculations, for both the primal and the differentiation. Symbolic transformation and analysis of Lines 6-7 in Figure~\ref{fig:bgdExample}(a) can show that the loop involves the calculations of $\sum_i x[i]*y[i]$ and $\sum_i x[i]*x[i]$; when the analysis scope spans across the entire \texttt{while} loop via $\phi$-calculus, the optimization can easily recognize that the two summations repeat in every iteration of the \texttt{while} loop and can be hoisted out of the \texttt{while} loop. Similar phenomena are in the differentiation. Neither the default optimizers in the Java Runtime underlying Kotlin or the coarsening without $\phi$-calculus can recognize and take advantage of that. (iii) It saves the remaining AD overhead that the version by coarsening without $\phi$-calculus has to suffer. 



\noindent{\bf Brachist.} 
This program calculates the Brachistochrone curve (i.e., curve of fastest descent), which is the one lying on the plane between a point A and a lower point B (called anchor points), where B is not directly below A, on which a bead slides frictionlessly under the influence of a uniform gravitational field to a given end point in the shortest time. In each iteration, the program computes the time taken by the bead to slide down the slope by summing the time it takes for each section of the current curve, and then gets the gradient of every section over the total time. The three configurations correspond to the number of sections that the target curve is regarded to be composed of. 
This is a relatively easy case, but it demonstrates an important scenario where coarsening can remove the entire primal computation. \rev{The entire primal code to compute the time taken to slide down the slope is identified as the SOI.} With coarsening, the AD tool can symbolically differentiate the entire computation. Because the program only needs the gradients to update the curve in each iteration, it can now forego the computation of the total time as the gradients can be directly computed. Therefore the coarsening optimization removes the entire primal computation, making the program's overall speedups even more than the speedups on the gradients calculations. As Table~\ref{tab:results} shows, the speedups on the differentiation part is modest (due to the simplicity and regularity of the code), but the end-to-end executions get more significant speedups (e.g., 1.8-2.38$\times$ versus 1.03-1.42$\times$ on macBook). The overall speedups are more pronounced on the smaller inputs because the primal computation weights more in those runs. 

\noindent{\bf CartPole.}
CartPole is a deep reinforcement learning program as already introduced in Section~\ref{sec:overview}. The three configurations corresponding to the number of exploration steps observed before learner updates the model parameters. It shows the least speedups among all the benchmarks, not because coarsening is not effective, but because the small portion of the optimized code weighs in the overall program. Recall in Figure~\ref{fig:cartpole}, the primal computation of CartPole contains two parts, the Neural Networks(NN) and the environment update. As CartPole uses a simple simulation environment, the environment update part weighs only about 15\% of the primal time, with the rest dominated by the NN. \rev{As the NN has a standard structure and the default gradient calculation is through a manually written highly polished vendor library rather than the AD, the SOI is the second part, which updates the environment. There are five $\phi$ functions in the SOI.} The speedups on the differentiation of the environment update part are actually significant:
\begin{quote}
\small
    \begin{tabular}{p{1.8in}|c|c|c|c|c|c}\hline
         &  \multicolumn{3}{c|}{devServer} & \multicolumn{3}{c}{macBook} \\\cline{2-7}
    Config   & 1 & 2 & 3 & 1 & 2 & 3 \\\hline
    Speedup(X) on differentiating the environment update part & 2.68 &	2.63&	3.19&	3.01&	2.73&	3.21\\\hline
\end{tabular}
\end{quote}
In cases where reinforcement learning is applied to more complex environments, the speedups on the end-to-end execution by coarsening are expected to be more substantial.

\begin{figure}
    \centering
    \includegraphics[width=.3\textwidth]{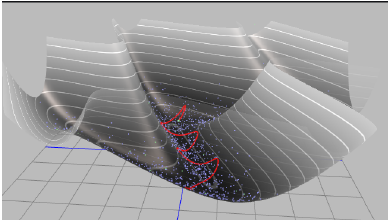} \\
    (a) HMC sampling in a space (\cite{hmcGraph}) \\[4mm]
    \begin{tabular}{cc}
    \includegraphics[width=.31\textwidth]{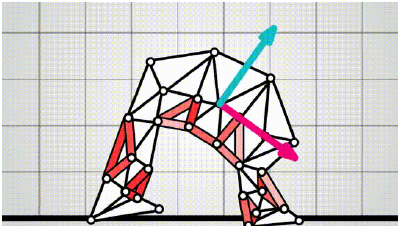} &
    \includegraphics[width=.32\textwidth]{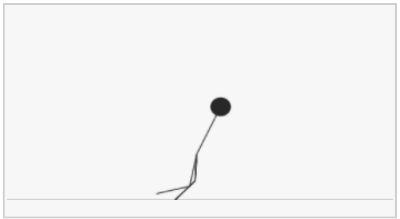}\\
    (b) HookeanSpring (\cite{rojas2019-drl-sbl}) & (c) QWOP
    \end{tabular}
    \caption{Illustrations of several benchmarks.}
    \label{fig:illustration}
\end{figure}

\noindent{\bf HMC.}
HMC stands for Hamiltonian Monte Carlo. It is one of the main algorithms in Probabilistic Programming or Statistics for finding out posterior distributions of random variables through a carefully designed Monte Carlo sampling process~\cite{neal2012mcmc}, as illustrated in Figure~\ref{fig:illustration}(a). 

We first gives a conceptual view of HMC. The core sampling function in HMC takes in two functions as part of the arguments: function "U" and function "grad\_U". The former does the primal computation and the latter computes the derivative of "U". HMC calls them many times on different values, more often on "grad\_U" than on "U". In this benchmark, HMC is used for logistic regression. Logistic regression is a classic method for classification. HMC is used to estimate the posterior distribution of the parameters in the logistic regression model. Its "U" function is as follows:
\[
U(\beta) = \beta^TX^T(y-1_n)-1^T_n[log(1+e^{-X\beta})]-\frac{\beta^T\beta}{2\sigma^2_\beta}
\]
\noindent where, $\beta$ are the parameters in logistic regression, $X$ and $y$ are the input and response data (training data), $\sigma^2_\beta$ is a hyperparameter (1000). All are vectors except that $X$ is a matrix.
The three configurations correspond to three different training data sets. 

A feature of HMC is that when it needs the gradient on some values, it often cares about the gradients but not the actual primal value. It does call the primal function at some places, but on values different from those needed for the calculation of the gradients.

But in the default implementation on AD, anytime gradient is needed, the primal is called because of the inherent requirement for AD to work as we have described in Section~\ref{sec:background}. \rev{Coarsening takes the entire primal function "U" as the SOI, and there are one two-level nested loop and another two single-level loops in the code with one of them containing an if-else statement.} Coarsening is able to symbolically differentiate it and generate the entire "grad\_U" function. It hence can save many primal computations. Overall, the speedups are 2.3-3.6$\times$. 

One special note on HMC is that the exponential term $e^{-X\beta}$ can easily result in value overflow; special treatments must be given to large exponents (e.g., using $-X\beta$ to approximate $log(1+e^{-X\beta}$ if $-X\beta$ exceeds 80). The default AD-based tool uses a masking function to deal with that (similar to "where" in PyTorch). Without the masking scheme, an alternative would be to wrap each element in $-X\beta$ in a Tensor to discern their following uses; which creates huge runtime overhead, making the program run about 25$\times$ slower. Without the dependence on Tensor or operator overloading, coarsening is not subject to the problem; when it generates the code, it directly generates the appropriate code for the cases where the exponential value is large.


\noindent{\bf HookeanSpring.}
HookeanSpring is a physical simulation program. It simulates mass-spring systems as illustrated in Figure~\ref{fig:illustration}~\cite{rojas2019-drl-sbl}. It demonstrates the transitions of physics-based states as energy minimization procedures. The program keeps optimizing the vertex positions of a spring system to find some configuration that minimizes the total elastic energy. Every spring has some preferred rest length and they naturally tend to recover their rest shapes over time. Each optimization step uses the gradients of the spring vertex locations regarding to the system energy. The three configurations correspond to three sizes of the spring system in terms of the number of spring vertices. \rev{Coarsening is able to take the entire energy calculation of the Spring system as the SOI and symbolically differentiate it.} As a result, the primal computation which computes the system energy can be completely removed. The speedups are 4--11$\times$. The program even runs faster than the original primal computation alone; the following table shows the times taken in one iteration of the simulation and the relative speedups (median values of repeated measurements are used):
\begin{quote}
\small
    \begin{tabular}{c|c|c|c|c|c|c}\hline
         &  \multicolumn{3}{c|}{devServer} & \multicolumn{3}{c}{macBook} \\\cline{2-7}
    Config   & 1 & 2 & 3 & 1 & 2 & 3 \\\hline
    Original primal only($\mu s$) &    49.19	& 51.92& 	114.20&	43.37&	47.01&	52.58\\
    Exec. after coarsening($\mu s$) & 14.49&	26.76&	51.49&	14.46&	21.61&	41.17\\
    Speedups($\times$) & 3.39&	1.94&	2.22&	3.00&	2.18&	1.28\\\hline
\end{tabular}
\end{quote}
This result is significant because there has been a common perception that a program would take a lot more time to run if automatic differentiation is added into it. For example, a previous work considers 2.4-4$\times$ slowdown after adding automatic differentiation as already close to the optimal~\cite{Hogan+:TMS2014}. This coarsening result shows that with coarsening, after adding automatic differentiation, a program can even run several times faster.  


\noindent{\bf QWOP.}
QWOP is an avatar motion optimization program. It trains a virtual stick figure to run as far as possible by providing a schedule for how much each muscle should be extended, as illustrated in Figure~\ref{fig:illustration}(c). The three configurations correspond to three configurations of the mass of the body parts of the stick figure. The special aspect about this program is that its core part is a 225-line function with 13 loops and many if-else statements. After loop unrolling, the function becomes 1117-line long. Coarsening can successfully deal with the function, getting two SOIs, and achieving 1.17-1.51$\times$ overall speedups. 

\rev{
\subsection{Potential on Other AD Tools}

Coarsening is a general optimization for AD. To check its potential benefits to AD tools beyond DiffKt, we examined the performance of the benchmark \texttt{BGDHyperOpt} on three other AD tools: JAX~\cite{jax2018github} for Python, Zygote~\cite{Innes:MLSys2020} for Julia, and Adept~\cite{Hogan+:TMS2014} for C++.

For each of the three AD tools, we have two versions of the benchmark BGDHyperOpt: (i) the baseline version which uses the default AD offered by the tool; (ii) the coarsened version optimized by coarsening. The latter was written based on the results from our symbolic engine. Table~\ref{tab:otherAD} reports the speedups of the coarsened versions over the baseline counterparts. Please note that the JAX baseline version already uses its JIT (the JIT gives 1.3-1.47X speedups over the default version that uses no JIT). The execution times were measured on the Macbook after warm-ups. The speedups are 66$\times$-335$\times$, even greater than on DiffKt, indicating the potential of coarsening as a general AD optimization technique.}

\begin{table}[h]
    \centering
    \caption{\rev{Speedups of the coarsened version over the baseline version on \texttt{BGDHhyperOpt}}}\label{tab:otherAD}
    \begin{tabular}{|c|c|c|}\hline
    {AD Tool (Language)}     & \multicolumn{2}{|c|}{{Input Size}} \\
    & {1000} & {2000}  \\\hline
    JAX (Python)~\cite{jax2018github}    & 87.4X & 335.1X \\
    Zygote (Julia)~\cite{Innes:MLSys2020}  & 150X & 90.8X \\
    Adept (C++)~\cite{Hogan+:TMS2014} & 66X & 96.2X \\\hline
    \end{tabular}
\end{table}



\section{Discussions}

The study has demonstrated the significant benefits of coarsening on the Kotlin AD tool on first-order backward AD, the most popular kind of AD. It is easy to see that the technique can help other types of AD implementations (e.g., forward or mixed directions and higher-order differentiation) as the outcome of coarsening can always be used as a shortcut on the AD chains. 

\rev{Our exploration of coarsening is at the static compilation time. The technique is potentially applicable at runtime as well, which could be especially meaningful for languages (e.g., Python) that are difficult for static time analysis and transformations. In that case, runtime profiling could be useful, and extra care (e.g., hot paths based selective optimizations) may be necessary to minimize the time overhead of symbolic manipulations.}

\rev{The current coarsening optimization applies to both regular and irregular loops as mentioned before. But there is code with unstructured control flows where it is even unclear what the loop is (e.g., code formed by go-to statements in certain languages). In those cases, the SOIs could be set to the sections within the branches that form the unstructured control flows.}

\rev{
With coarsening, the compilation time does not have noticeable changes except for the time taken by the symbolic engine in doing symbolic differentiation and other symbolic manipulations. As mentioned, as a proof of concept, the current implementation uses an extended Sympy for that. Written in Python, Sympy is not the most efficient symbolic engine. For the benchmarks in the experiment, it takes up to a minute to do symbolic differentiation.
}

Coarsening is based on the SSA form of a program. When a program has assignments to arrays, array SSA~\cite{Knobe+:POPL98} would be necessary to discern the different ranges of data elements in an array when they are treated differently in the program. The representation and corresponding analysis are more complicated than on the basic SSA form. We found that for AD programs written in Tensor-based AD libraries, in most cases, array SSA is not necessary. It is because in those programs, if there are large arrays, operations on them are typically written as Tensor operations (e.g., $C=A+B$ for \texttt{Loop: c[i] = a[i]*b[i]}) without explicit references to individual array elements; for such representations, the standard SSA still applies. Even if sometimes a part of the array elements are treated differently from others, Tensor operations still suffice via Tensor masking operations (e.g., the \texttt{HMC} case). In the cases where individual array elements are used and updated differently, those arrays are usually short and are used in small loops; loop unrolling and scalar conversion can easily turn the code into a form amenable for the standard SSA. Nonetheless, integration of array SSA with coarsening could still be useful especially for AD tools without Tensor-like abstractions. 


\section{Related Work}

There is a large body of work on efficient AD. Optimizations range from checkpointing~\cite{Dauvergne+:2006} to edge/vertex eliminations on computation graphs~\cite{Dixon1991}, combination of forward and backward differentiation~\cite{Baydin+:JML2018}, loop transformations~\cite{Shaikhha+:ICFP2019}, and so on.
A recent work~\cite{Sherman+:POPL2021} proposes a differentiable programming language to deliver a semantics for higher-order functions, higher-order derivatives, and Lipschitz but non-differentiable functions. Some of the relevant studies have been mentioned in earlier sections, and more on AD for machine learning can be seen in recent surveys~\cite{Baydin+:JML2018,vanmerrienboer2019automatic, Margossian:ADSurvey2019}. Coarsening can be regarded as an optimization complementary to those existing AD optimizations: They can be used together, with coarsening offering shortcuts and the other optimizations improving the remaining AD operations.

There are many tools capable of doing symbolic differentiation. Examples include the Calculus module in Julia~\cite{calculusJulia}, SageMath~\cite{sagemath}, KotlinGrad~\cite{considine2019kotlingrad}, and Acumen which maps from analytical models to simulation codes via symbolic differentiation~\cite{Zhu+:ICPS2010}. None of them have addressed the complexities from control flow on symbolic differentiation, or the systematic integration of symbolic differentiation with AD. 
\rev{The existing symbolic engines can differentiate only expressions not programs. For cases with simple control flows where the problem of interest involves only several conditional cases,  the user could enumerate those cases and use the existing symbolic engines to differentiate them each. That practice does not apply to code with loops or many branches. The $\phi$-calculus in this work offers a solution to the complexity.}

Several recent studies have challenged the common criticisms of ``expression swell'' of symbolic differentiation~\cite{wang2019demystifying,laue2020equivalence}. Even though the arguments may differ in form, the main points are similar: If placeholders are used to store intermediate differentiation results for reuses, the problem can be largely alleviated. The design of our reuse-aware SOI identification in coarsening is based on a similar insight, but provides a systematic way to deal with the tradeoff between reuse and the granularity of symbolic differentiation.

Expression templates have been used in both forward and backward AD implementations to reduce runtime space and time overhead\cite{Aubert+:CVS2001,Phipps+:ExpTemp2012,Sagebaum+:TMS2017}. In Adept, for instance, during the primal computations, the algorithm records backward operations onto a stack. Its use of expression templates in C++ helps avoid invocations of virtual function calls at runtime, and hence reduces the amount of objects needed to allocate to hold intermediate results. Differentiation happens however still at each individual operation. There is no symbolic differentiation or symbolic simplifications or optimizations in a large scope. Moreover, as with all other operator overloading based AD, these solutions also require primal computations to be executed before gradients can be computed. As a result, the highly optimized implementations are still 2.4-4X slower than the original algorithm (without gradients calculations)~\cite{Hogan+:TMS2014}. Coarsening optimization, in comparison, harnesses large-scope optimization opportunities, and can sometimes forego the primal computations completely, yielding even a higher speed than the original algorithm has as Section~\ref{sec:evaluation} has shown.

Since it was first proposed in late 1980s~\cite{Cytron+:POPL1989}, SSA has been widely adopted in program representations in compilers. Gated SSA (GSA) proposes to explicitly specify the conditions in the $\phi$ functions, and introduces notations to distinguish loop entry, loop exit, and normal $\phi$-functions~\cite{Ottenstein+:PLDI1990}, which share some similarities with part of the $\phi$-notations in this work. The loop notations in $\phi$-calculus is inspired by the notations in Glore~\cite{Ding+:OOPSLA2017}, a work that detects large-scoped loop invariants. In 1990s and early 2000s, there were a number of papers on recognizing and substituting inductive variables in loops based on SSA through symbolic analysis~\cite{Tu+:ICS1995}. An example is the symbolic analysis based on Chains of Recurrences~\cite{Engelen:CC2001,Engelen+:ICS2004}. The purpose is to convert the array subscripts in a loop into a form ready for parallelization-oriented dependence analysis. For symbolic differentiation, what is needed is not only symbolic treatment to inductive variables but derivations of the closed-form expressions for all the computations that are related with the active variables, hence the need for the $\phi$-calculus. 
As a type of standard IR, SSA is also the IR leveraged in recent AD compilers, such as the Zygote for Julia~\cite{innes2019dont,Innes:MLSys2020}, which is a pure AD tool without systematic integration of symbolic differentiation. 

Besides symbolic and algorithmic differentiation, there is another approach called numerical differentiation, which uses finite difference approximations. But because it is inaccurate and scales poorly for gradients, it is rarely used for machine learning where gradients with respect to millions of parameters are common.

\section{Conclusion}

This paper has presented {\em coarsening}, a novel optimization that expands the scope of symbolic differentiation and systematically integrates symbolic differentiation with AD. It builds on two key innovations: the {\em $\phi$-calculus} and the {\em reuse-aware SOI identification}. The {\em $\phi$-calculus} offers the first mechanism that allows symbolic differentiation to apply on code with complicated control flow, while the {\em reuse-aware SOI identification} provides an algorithm to deal with the tension between computation reuse and coarsening. Experiments on several AD tools and various settings demonstrate that coarsening is an effective optimization for AD. It can remove the overloading overhead in AD and at the same time harness the benefits of symbolic optimizations and differentiation, yielding several times to two orders of magnitude speedups. 

\bibliographystyle{abbrv}
\bibliography{all}

\begin{thebibliography}{10}
\bibitem{vanmerrienboer2019automatic}
B.~van Merri{\"{e}}nboer, O.~Breuleux, A.~Bergeron, and P.~Lamblin.
\newblock Automatic differentiation in {ML:} where we are and where we should
  be going.
\newblock {\em CoRR}, abs/1810.11530, 2018.
\end{thebibliography}


\end{document}